\documentclass[a4paper,11pt]{article}
\pdfoutput=1 

\usepackage{jheppub} 

\usepackage[T1]{fontenc} 
\usepackage{comment}
\usepackage{tikz}
\usepackage{amsmath,amssymb}

\title{Invariant traces of the flat space chiral higher-spin algebra as scattering amplitudes}

\author{Dmitry Ponomarev}

\affiliation{Institute for Theoretical and Mathematical Physics,\\
Lomonosov Moscow State University, Lomonosovsky avenue, Moscow, 119991, Russia}
\affiliation{I.E. Tamm Theory Department, Lebedev Physical Institute,\\
 Leninsky avenue, Moscow, 119991, Russia}

\emailAdd{ponomarev@lpi.ru}

\abstract{
We sum up two- and three-point amplitudes in the chiral higher-spin theory over helicities and find that these quite manifestly have the form of invariant traces of the flat space chiral higher-spin algebra. We consider invariant traces of products of higher numbers of on-shell higher-spin fields and interpret these as higher-point scattering amplitudes. This construction closely mimics its anti-de Sitter space counterpart, which was considered some time ago and was confirmed holographically.}

\begin{document} 
\maketitle
\flushbottom

\section{Introduction}

Interactions of higher-spin gauge fields in flat space are severely constrained by no-go theorems. Starting from rather natural and general sets of assumptions, these prove that higher-spin gauge fields cannot interact in the Minkowski space, see  \cite{Weinberg:1964ew,Coleman:1967ad} for earlier no-go theorems and \cite{Bekaert:2010hw} for a comprehensive review on the topic. Despite that, based on the results \cite{Metsaev:1991mt,Metsaev:1991nb}, the chiral higher-spin theory was put forward in \cite{Ponomarev:2016lrm}.
This theory is formulated in the light-cone gauge, it involves only cubic vertices and can be regarded as a higher-spin generalisation of self-dual Yang-Mills and self-dual gravity \cite{Ponomarev:2017nrr}. Similarly to the latter theories, the chiral higher-spin theory is integrable. Due to that its only non-trivial amplitude is the three-point one, which is, moreover, non-trivial only for complex momenta, while for real momenta it trivialises  due to peculiarities of massless three-point kinematics.
These expectations based on integrability were confirmed by explicit computations  and, moreover, some results are available at the quantum level \cite{Skvortsov:2018jea,Skvortsov:2020wtf,Skvortsov:2020gpn}.
 Thus, despite the chiral higher-spin theory is given by a non-linear action it gives rise to, in effect, trivial scattering in complete agreement with the no-go theorems.

At the same time, existence of higher-spin theories in the anti de-Sitter space is strongly supported by holography \cite{Sezgin:2002rt,Klebanov:2002ja}. In the case of unbroken higher-spin symmetry the dual conformal field theories are free \cite{Maldacena:2011jn,Boulanger:2013zza,Alba:2013yda}. Accordingly, the conformal field theory correlators reinterpreted as bulk higher-spin scattering amplitudes are rather degenerate -- in particular, in the Mellin representation the associated amplitudes are given by distributions  \cite{Taronna:2016ats,Bekaert:2016ezc}, -- but still non-vanishing. It seems natural to expect that a similar pattern may hold for higher-spin theories in flat space as well -- namely, that higher-spin scattering in flat space can be made non-trivial beyond three-point functions, though, it should be still rather singular not to contradict the no-go theorems.

To construct these higher-point amplitudes we will use symmetries as a main guiding principle. In the AdS case, higher-spin symmetries \cite{Fradkin:1986ka,Eastwood:2002su,Vasiliev:2003ev} turn out to be strong enough to fix scattering amplitudes up to an independent overall factor for each $n$-point amplitude.
A particularly efficient approach to implement this idea in AdS${}_4$ is to use the spinorial realisation of the higher-spin algebra \cite{Fradkin:1986ka} and 
to construct  amplitudes as invariant traces \cite{Vasiliev:1986qx} of this algebra. The key advantage of this approach is that it makes the higher-spin symmetry of amplitudes manifest.
The relevant analysis was carried out in \cite{Colombo:2012jx,Didenko:2012tv,Gelfond:2013xt,Didenko:2013bj}, moreover, the resulting amplitudes were found to give rise to correct correlators in the holographically dual theory. The approach based on invariant traces has then been extended to other setups, in particular, to theories with slightly broken \cite{Maldacena:2012sf} higher-spin symmetry \cite{Gerasimenko:2021sxj}.

In the present paper we sum up two- and three-point functions of the chiral higher-spin theory over helicities employing the spinor-helicity formalism, see \cite{Elvang:2013cua} for review. Besides having other advantages, which make it an indispensable tool for computing amplitudes in massless theories, in the higher-spin context the spinor-helicity formalism is beneficial as it allows one to adopt the techniques based on $sl(2,\mathbb{C})$ spinors to the four-dimensional flat-space case \cite{Bolotin:1999fa,Nagaraj:2018nxq,Nagaraj:2019zmk,Nagaraj:2020sji}.   
Having summed the amplitudes over helicities -- which required a certain well-known regularisation -- we find that the results quite manifestly have the form of invariant traces of products of higher-spin fields\footnote{For a comprehensive discussion of various associative products and their traces in the higher-spin context see \cite{Joung:2014qya}.}, in close analogy with the associated AdS space results \cite{Colombo:2012jx,Didenko:2012tv,Gelfond:2013xt,Didenko:2013bj}. The algebra that we thus obtain has already appeared in different forms in \cite{Ponomarev:2017nrr,Krasnov:2021nsq,Krasnov:2021cva,Skvortsov:2022syz,Sharapov:2022dfw}. 
By employing the associative product and the cyclic trace that we extracted from the low-point amplitudes, we construct invariant traces of higher numbers of on-shell higher-spin gauge  fields, which are then naturally interpreted as scattering amplitudes. 

This paper is organised as follows. We start by reviewing the relevant preliminaries of the spinor-helicity formalism in four-dimensional  flat space. Then, in section \ref{sec:3} we sum the two-point amplitude of higher-spin fields in the spinor-helicity formalism over helicities. Next, we evaluate the analogous sum for three-point amplitudes in the chiral higher-spin theory in section \ref{sec:4}.  In the following section we show how these amplitudes allow one to define the associative product of the flat space higher-spin algebra and the associated cyclic trace.
To highlight similarities of these results with their AdS counterpart, we review the construction of  higher-spin invariant AdS amplitudes in section \ref{sec:6}. Then, in section \ref{sec:7} we briefly discuss various properties of these amplitudes and, finally, we conclude in section \ref{sec:8}. This paper has a number of appendices, which collect some technical results.

\section{Preliminaries and conventions}

We start by reviewing our conventions and basic results on the spinor-helicity formalism for massless fields in the 4d Minkowski space. For a more comprehensive review we refer the reader to  \cite{Elvang:2013cua}.

We use the Minkowski space metric in the mostly plus signature $\eta={\rm diag}(-,+,+,+)$.
The Pauli matrices are defined by
\begin{equation}
\label{3feb1x1}
\sigma^0 = 
\left(\begin{array}{cccc}
1 &&& 0\\
0 && &1
\end{array}\right), \quad \sigma^1 = 
\left(\begin{array}{cccc}
0 & && 1\\
1& && 0
\end{array}\right), 
\quad 
 \sigma^2 = 
\left(\begin{array}{ccc}
0 && -i\\
i&& 0
\end{array}\right), \quad
 \sigma^3 = 
\left(\begin{array}{ccc}
1 && 0\\
0&& -1
\end{array}\right).
\end{equation}
These allow us to convert a Lorentz vector to a $sl(2,\mathbb{C})$ bispinor 
\begin{equation}
\label{25jan1}
p_{\alpha\dot \alpha}\equiv p_\mu(\sigma^\mu)_{\alpha\dot \alpha}= \left(
\begin{array}{cc}
-p^0+p^3 & p^1- ip^2\\
p^1+ip^2 & -p^0-p^3
\end{array}
 \right),
\end{equation}
while the inverse formula reads
\begin{equation}
\label{16may1}
p_a = -\frac{1}{2}(\sigma_a)^{\dot\alpha\alpha}p_{\alpha\dot\alpha}.
\end{equation}
The $sl(2,\mathbb{C})$ indices are raised and lowered with the Levi-Civita tensor according to the convention
\begin{equation}
\label{c16may1}
\lambda^\alpha = \epsilon^{\alpha\beta} \lambda_{\beta}, \qquad \lambda_{\beta}=\epsilon_{\beta\gamma}\lambda^\gamma,
\qquad
\bar\lambda^{\dot\alpha} = \epsilon^{\dot\alpha\dot\beta} \bar\lambda_{\dot\beta}, \qquad \bar\lambda_{\dot\beta}=\epsilon_{\dot\beta\dot\gamma}\bar\lambda^{\dot\gamma},
\end{equation}
where
\begin{equation}
\label{c16may2}
\epsilon^{\alpha\beta}=\epsilon^{\dot\alpha\dot\beta}
=\left(\begin{array}{ccc}
0 && 1\\
-1&& 0
\end{array}\right)=\epsilon_{\alpha\beta}=-\epsilon_{\dot\alpha\dot\beta}.
\end{equation}
For spinor products we use 
\begin{equation}
\label{16mayx1}
\langle ij\rangle = \lambda^i_\alpha \lambda^{j\alpha} = \lambda^i_{\alpha} \lambda^j_{\beta}\epsilon^{\alpha\beta}, \qquad 
[ ij]= \bar\lambda^i_{\dot\alpha} \bar\lambda^{j\dot\alpha} = \bar\lambda^i_{\dot\alpha} \bar\lambda^j_{\dot\beta}\epsilon^{\dot\alpha\dot\beta}.
\end{equation}

For light-like momenta $p^ap_a=0$, one has ${\rm det}(p_{\alpha\dot\alpha})=0$, so the momentum bispinor can be factorised
\begin{equation}
\label{27jan1}
p_{\alpha \dot\alpha}=-\lambda_\alpha \bar\lambda_{\dot \alpha}. 
\end{equation}
For real momenta and positive energy, $p^0>0$, it can be solved for $\lambda$ and $\bar\lambda$ with $\bar\lambda=(\lambda)^*$. Explicitly, one has
\begin{equation}
\label{27jan3}
\lambda_\alpha =
\left(
\begin{array}{c}
\sqrt{p^0-p^3}\\
-\frac{p^1+ip^2}{\sqrt{p^0-p^3}}
\end{array}
\right), \qquad 
\bar\lambda_{\dot \alpha} =
\left(
\begin{array}{c}
\sqrt{p^0-p^3}\\
-\frac{p^1-ip^2}{\sqrt{p^0-p^3}}
\end{array}
\right).
\end{equation}
For negative energies we will use $\bar\lambda=-(\lambda)^*$. Accordingly, $\lambda \to -\lambda$ or $\bar\lambda\to-\bar\lambda$ act as $p\to -p$ and, thereby, exchange positive and negative energy modes. In the following,  we will assume that momenta are complex, therefore, $\lambda$ and $\bar\lambda$ are independent. 

In terms of $sl(2,\mathbb{C})$ spinors, the Poincare algebra can be realised as
\begin{equation}
\label{16may2}
\begin{split}
J_{\alpha\beta}&=i\left(\lambda_\alpha\frac{\partial}{\partial \lambda^\beta}+ \lambda_\beta\frac{\partial}{\partial \lambda^\alpha}\right),\\
\bar J_{\alpha\beta}&=i\left(\bar \lambda_{\dot\alpha}\frac{\partial}{\partial \bar\lambda^{\dot\beta}}+ \bar\lambda_{\dot\beta}\frac{\partial}{\partial \bar\lambda^{\dot\alpha}}\right),\\
P_{\alpha\dot\alpha}&=-\lambda_{\alpha}\bar\lambda_{\dot\alpha},
\end{split}
\end{equation}
where the operators are assumed to be acting on functions $\Phi(\lambda,\bar\lambda)$ on $\mathbb{C}^2/\{ 0\}$.
All operators in representation (\ref{16may2}) commute with
\begin{equation}
\label{16may3}
H\equiv \frac{1}{2} \left(\bar N- N\right),
\qquad \bar N  \equiv \bar\lambda^{\dot\alpha}\frac{\partial}{\partial\bar\lambda^{\dot\alpha}} , \qquad N\equiv \lambda^\alpha \frac{\partial}{\partial \lambda^\alpha} 
\end{equation}
where $H$ is called the helicity operator. By requiring that the helicity operator takes  definite value 
\begin{equation}
\label{16may4}
H\Phi_h = h\Phi_h
\end{equation}
representation (\ref{16may2}) becomes irreducible. It describes positive energy modes of massless fields of helicity $h$. It is required that  $h$ is  half-integer, otherwise, 
$\Phi(\lambda,\bar\lambda)$ is not single-valued. 
In what follows, we will discuss bosonic fields only. These are fields of integer helicity, which is equivalent to 
\begin{equation}
\label{16may5}
\Phi(\lambda,\bar\lambda) = \Phi (-\lambda,-\bar\lambda).
\end{equation}
Besides that, real fields in the coordinate representation correspond to 
\begin{equation}
\label{16may5x1}
\Big(\Phi(\lambda,\bar\lambda)\Big)^* = \Phi (-\lambda,\bar\lambda),
\end{equation}
that is positive and negative energy modes in the Fourier space have complex conjugate coefficients.

Amplitudes transform as forms on wave functions. This means that an $n$-point amplitude on each leg transforms as $A(\lambda,\bar\lambda)$, with the transformation law defined so that the natural inner product
\begin{equation}
\label{16may6}
\int d^2\lambda d^2\bar\lambda A(\lambda,\bar\lambda) \Phi(\lambda,\bar\lambda)
\end{equation}
remains invariant. As it is not hard to see, this implies that  the Poincare algebra is realised on amplitudes  in the same way as on wave functions,  (\ref{16may2}), except for the replacement $P\to -P$. Besides that, a wave function of helicity $h$ non-trivially couples via (\ref{16may6}) to an amplitude with $HA = -h A$.

In addition to the extensive use of $sl(2,\mathbb{C})$ spinors, which enables one to efficiently solve the massless on-shell condition $p^2=0$, another inherent feature of the spinor-helicity formalism is the particular form of polarisation vectors for potentials. For example, the massless vector field  has two independent degrees of freedom on-shell, which carry helicities $+1$ and $-1$. In the spinor-helicity formalism the associated polarisation vectors $\varepsilon^+$ and $\varepsilon^-$ are taken in the form
\begin{equation}
\label{9may6}
\varepsilon_\mu^+ = \frac{1}{\sqrt{2}}\frac{(\sigma_\mu)^{\alpha\dot\alpha}\bar\lambda_{\dot\alpha}\mu_{\alpha}}{\mu^\beta \lambda_\beta}, \qquad \varepsilon_\mu^- = (\varepsilon_\mu^+)^*.
\end{equation}
These are normalised so that $\varepsilon^+\cdot\varepsilon^-=1$, while other products vanish. In (\ref{9may6}) $\mu$ is an arbitrary auxiliary spinor called the reference spinor. Changes in $\mu$ correspond to gauge transformations. Reference spinors $\mu$ and $\bar\mu$ together define a null vector according to the standard vector-spinor dictionary (\ref{16may1}). Potentials for higher-helicity fields are built in terms of  products of polarisation vectors (\ref{9may6}). For example, for positive integer helicity-$h$ field, the polarisation tensor for the potential is given by $(\varepsilon_\mu^+)^h$.

\section{Two-point amplitude}
\label{sec:3}

In the present section we compute the two-point amplitude of the chiral higher-spin theory in the spinor-helicity formalism by simply summing the canonically normalised two-point amplitudes for individual fields over helicities. It is not typical to speak of two-point amplitudes as these do not correspond to any non-trivial scattering. However, in the present context we are primarily interested in the transformation properties under the action of the Poincare algebra and, in this sense, the two-point Wightman functions will be taken as two-point amplitudes, since the former transform as two-forms on the on-shell fields.

To start, we consider the two-point amplitude of the massless scalar field and translate it to the spinor-helicity representation.
The canonically normalised scalar two-point amplitude reads\footnote{We will often integrate amplitudes against wave functions. In this case, integrated amplitudes will be denoted by $G$, while amplitudes themselves will be denoted $A$. Note also that we include the momentum-conserving delta function into the amplitude.}
\begin{equation}
\label{9may1}
G_2^0 =\int d^4p_1d^4p_2\theta(p_1^0) \delta(p_1^2)\delta^4 (p_1+p_2)\Phi_1(p_1)\Phi_2(p_2).
\end{equation}
To convert it to the spinor-helicity formalism, we first need to isolate on-shell delta functions, associated with each external line, which will then be dropped, as $sl(2,\mathbb{C})$ spinors solve the mass-shell condition identically. This can be done as follows
\begin{equation}
\label{9may2}
\begin{split}
G_2^0 &= \int d^3\vec p_1d^3\vec p_2 \frac{1}{2| \vec p_1|}
\delta^3 (\vec p_1+\vec p_2)\Phi_1(|\vec p_1|,\vec p_1)\Phi_2(-|\vec p_2|,\vec p_2)\\
&=\int d^4p_1d^4p_2\theta(p_1^0) \delta(p_1^2)\theta(-p_2^0) \delta(p_2^2) 2|\vec p_1|\delta^3 (\vec p_1+\vec p_2)\Phi_1(p_1)\Phi_2(p_2),
\end{split}
\end{equation}
where in the first line, the first argument of $\Phi$ is energy, while the second one is spatial momentum.
The kernel of this integral with the on-shell delta functions dropped can be identified with the two-point amplitude
\begin{equation}
\label{9may3}
A_2= 2|\vec p_1|\delta^3 (\vec p_1+\vec p_2).
\end{equation}

We would like to emphasise that (\ref{9may3}) is not manifestly Lorentz invariant. This happens due to the fact that for on-shell momenta spatial momentum conservation implies conservation of energy, so the manifestly Lorentz covariant momentum-conserving $\delta^4(p_1+p_2)$ is singular on-shell. In the above manipulations we avoided this singularity by imposing momentum conservation only on spatial momenta, which lead to the loss of manifest Lorentz invaraince. 

The two-point amplitude (\ref{9may3}) can be readily converted to the spinor-helicity variables by employing the usual vector-to-spinor dictionary (\ref{16may1}), (\ref{27jan1}). 
Instead, we would like to bring (\ref{9may3}) to a somewhat different form, which seems to be more natural in the spinor-helicity context. More precisely, (\ref{9may3}) can be rewritten as
\begin{equation}
\label{28apr12}
A^0_2 = 4 \langle 1\mu \rangle [\mu 1] \delta(\langle 1\mu\rangle [\mu 1]+\langle 2\mu \rangle [\mu 2]) \delta(\langle 12 \rangle)\delta([12]),
\end{equation}
where $\mu$ and $\bar\mu$ are the standard reference spinors. 

To see that (\ref{28apr12}) is equivalent to (\ref{9may3}), we note that $\langle 12\rangle=0$ and $[12]=0$ imply that $p_1$ and $p_2$ are proportional to each other. Moreover, 
\begin{equation}
\label{9may4}
\langle 1\mu\rangle [\mu 1]+\langle 2\mu \rangle [\mu 2]=0
\end{equation}
entails that projections of $p_1$ and $p_2$ onto the vector associated with $\mu$ and $\bar \mu$ are opposite. Altogether, this implies $p_1+p_2=0$, as required\footnote{Strictly speaking, $p_1+p_2$ can still have a component along $\langle\mu\sigma\bar\mu]$. To avoid this, instead of  $\langle\mu\sigma\bar\mu]$ one can consider a time-like auxiliary vector $v$. Then, 
$v\cdot (p_1+p_2)=0$ does imply $p_1+p_2=0$. Still we will use  $\langle\mu\sigma\bar\mu]$ as the auxiliary vector as it is already present in the spinor-helicity formalism, together with the aforementioned subtlety, which manifests itself in $\langle \mu\lambda\rangle$ and $[\mu\lambda]$ appearing in denominators.}. The prefactor $\langle 1\mu \rangle [\mu 1]$ is introduced to make (\ref{28apr12}) of homogeneity degree zero in both $\mu$ and $\bar\mu$. As it is not hard to see, this entails that (\ref{28apr12}) does not, actually, depend on $\mu$ and $\bar\mu$ and, as a result, the two-point function is Lorentz invariant. Finally, the factor of four in (\ref{28apr12}) can be extracted by directly computing the Jacobian between the arguments of delta-functions in (\ref{28apr12}) and (\ref{9may3}) and matching these expressions. This is done in appendix \ref{app:jac}.

Having settled with the scalar two-point function, we proceed to the spin one case. The canonically normalised spin one two-point function is just the scalar one (\ref{9may2}), which is additionally supplemented with $\eta_{\mu\nu}$ that contracts Lorentz indices of on-shell potentials
\begin{equation}
\label{9may5}
\begin{split}
G_2^0 =\int d^4p_1d^4p_2\theta(p_1^0) \delta(p_1^2)\theta(-p_2^0) \delta(p_2^2) 2|\vec p_1|\delta^3 (\vec p_1+\vec p_2)\eta_{\mu\nu}\Phi^\mu(p_1)\Phi^\nu(p_2).
\end{split}
\end{equation}
Contracting the polarisation vectors for two fields in the case when the first field is of positive helicity, we obtain
\begin{equation}
\label{4may7x1}
\epsilon_1^+ \cdot \epsilon_2^-=-\frac{[1\mu]\langle \mu 2 \rangle}{[2\mu]\langle \mu 1\rangle}.
\end{equation}
Here it was taken into account that one of the fields has negative energy, which in the spinor-helicity formalism leads to the replacement $\lambda \to -\lambda$ and changes the sign of the polarisation vectors, see (\ref{9may6}). Combining this factor with the remaining kinematic factor, which is identical to that of the scalar field, we find 
\begin{equation}
\label{9may7}
A^1_2 = 4\left( -\frac{[1\mu]\langle \mu 2 \rangle}{[2\mu]\langle \mu 1\rangle} \right) \langle 1\mu \rangle [\mu 1] \delta(\langle 1\mu\rangle [\mu 1]+\langle 2\mu \rangle [\mu 2]) \delta(\langle 12 \rangle)\delta([12]).
\end{equation}
In a similar manner we can construct two-point functions for fields of any helicity. Eventually, we find that the two-point function for the helicity-$h$ field is given by 
\begin{equation}
\label{9may8}
A^h_2 = 4\left( -\frac{[1\mu]\langle \mu 2 \rangle}{[2\mu]\langle \mu 1\rangle} \right)^h \langle 1\mu \rangle [\mu 1] \delta(\langle 1\mu\rangle [\mu 1]+\langle 2\mu \rangle [\mu 2]) \delta(\langle 12 \rangle)\delta([12]).
\end{equation}

As the last step, we would like to sum the two-point amplitude over all integer helicities. This leads to the total two-point amplitude
\begin{equation}
\label{4may8}
A_2 = 4 \sum_{h=-\infty}^\infty \left(-\frac{\langle 1\mu\rangle [\mu 2]}{\langle 2\mu\rangle [\mu 1]} \right)^h  \langle 1\mu \rangle [\mu 1] \delta(\langle 1\mu\rangle [\mu 1]+\langle 2\mu \rangle [\mu 2]) \delta(\langle 12 \rangle)\delta([12]).
\end{equation}
To carry out this sum, we use 
\begin{equation}
\label{4may10}
\sum_{h=-\infty}^\infty z^h = \delta (1-z),
\end{equation}
which is the standard formula, see  e.g. \cite{VOA1,Nozaradan:2008zq} for applications in the vertex operator algebra literature\footnote{Summation in (\ref{4may10}) is done as follows. One should treat $\sum z^h$ as a distribution integrated against a test function $f(z)$ along the contour that encircles the origin in the complex plane. Then, by the Cauchy theorem, this integral reduces to $\sum a_h$, where $a_h$ are the Laurent series coefficients for $f(z)$. Assuming that $f(z)$ satisfies some properties -- e.g. that its Laurent series is bounded from both sides -- that ensure that $\sum a_h$ converges, we find that $\sum a_h=f(1)$, which allows us to infer (\ref{4may10}).}. This leads to
\begin{equation}
\label{4may11}
\begin{split}
A_2
=4\delta\left(\langle 2\mu\rangle [\mu 1]+ {\langle 1\mu\rangle [\mu 2]} \right) \langle 2\mu\rangle [\mu 1] \langle 1\mu \rangle [\mu 1] \delta(\langle 1\mu\rangle [\mu 1]+\langle 2\mu \rangle [\mu 2]) \delta(\langle 12 \rangle)\delta([12]).
\end{split}
\end{equation}

To simplify (\ref{4may11}), we recall that due to the vanishing of $\langle 12\rangle$ and $[12]$, one has 
\begin{equation}
\label{4may12}
\lambda_1 = \alpha \lambda_2, \qquad \bar\lambda_1 =\beta\bar\lambda_2.
\end{equation}
The remaining two delta functions imply
\begin{equation}
\label{4may13}
\alpha+\beta=0, \qquad \alpha\beta+1=0.
\end{equation}
Therefore, we find that the support of delta functions in (\ref{4may11}) consists of
\begin{equation}
\label{4may14}
\lambda_1 = \lambda_2, \qquad \bar\lambda_1=-\bar\lambda_2,
\end{equation}
and 
\begin{equation}
\label{4may15}
\lambda_1 = -\lambda_2, \qquad \bar\lambda_1=\bar\lambda_2.
\end{equation}
This entails that $A_2$ is, schematically, of the form
\begin{equation}
\label{9may9}
A_2
= (\dots) \delta^2(\lambda_1-\lambda_2)\delta^2(\bar\lambda_1+\bar\lambda_2)+
(\dots) \delta^2(\lambda_1+\lambda_2)\delta^2(\bar\lambda_1-\bar\lambda_2).
\end{equation}

Keeping in mind that we are dealing with bosonic fields, for which (\ref{16may5}) holds, we conclude that the two delta functions in (\ref{9may9}) are, actually, equal when integrated against the wave functions. By directly computing the Jacobian between the arguments of delta functions in (\ref{4may11}) and (\ref{9may9}),
-- which is done in appendix \ref{app:jac} --
 we find for the latter form of the amplitude the following result
\begin{equation}
\label{4may19}
 A_2 = 4  \delta^2(\lambda_1-\lambda_2)\delta^2(\bar\lambda_1+\bar\lambda_2).
\end{equation}
The extra factor of four in (\ref{4may19}) originates from the fact that 
each on-shell momentum $p^2=0$ is represented by two pairs of spinors: $(\lambda,\bar\lambda)$ and $(-\lambda,-\bar\lambda)$. Because of that the procedure that we followed above to convert two-point functions from the standard representation to the spinor-helicity one, leads to  overcounting with a factor of four -- two for each external line. Removing this factor, we find just
\begin{equation}
\label{9may10}
 A_2 =   \delta^2(\lambda_1-\lambda_2)\delta^2(\bar\lambda_1+\bar\lambda_2).
\end{equation}

Before finishing this section we would like to make a general comment on the manifest Lorentz covariance. As we noted before, the two-point amplitudes for individual spins (\ref{9may8}) are not manifestly Lorentz covariant. Moreover, by exploring all  possibilities to construct manifestly Lorentz covariant expressions, one can see that individual two-point amplitudes cannot be written in the manifestly covariant form at all. A similar phenomenon occurs for some of the three-point spinor-helicity amplitudes, which cannot be written in terms of contractions of Lorentz tensors. This illustrates -- what seems to be a persistent phenomenon, -- that by insisting on a particular representation of amplitudes or even on their manifest Lorentz covariance, we may overlook some of them.  
 We believe it is important to keep these examples in mind, as this phenomenon may be relevant for other problems. Let us note, that the total two-point function (\ref{9may10}), nevertheless, turns out to be manifestly Lorentz covariant.

\section{Three-point amplitude}
\label{sec:4}

 In the chiral higher-spin theory three-point amplitudes for individual helicities are given by
\begin{equation}
\label{9may11}
A_3^{h_1,h_2,h_3} = g\frac{\ell^{h-1}}{(h-1)!}[12]^{h_1+h_2-h_3} [23]^{h_2+h_3-h_1}[31]^{h_3+h_1-h_2}\delta^4(\lambda_1\bar\lambda_1+
\lambda_2\bar\lambda_2+\lambda_3\bar\lambda_3),
\end{equation}
which are non-vanishing only for $h\equiv h_1+h_2+h_3\ge 1$. Here $\ell$ and $g$ are two independent coupling constants. In the color-free case only amplitudes with $h$ even are non-trivial, though, (\ref{9may11}) is also consistent for matrix-valued fields. The amplitude (\ref{9may11}) admits interesting contractions, e.g. $\ell\to 0$, which can be then followed by truncations of the spectrum. Below, we will focus on the general case.

We will start by summing (\ref{9may11}) over helicities. Employing regularisation (\ref{4may10}), we find 
\begin{equation}
\label{16jul2}
\begin{split}
&\sum_{h_i}^\infty\frac{\ell^{h-1}}{(h-1)!}[12]^{h_1+h_2-h_3} [23]^{h_2+h_3-h_1}[31]^{h_3+h_1-h_2}\\
&\qquad=\sum_{h =1}^\infty\sum_{h_1=-\infty}^{\infty}\sum_{h_2=-\infty}^{\infty}\frac{\ell^{h-1}}{(h-1)!}
\left(\frac{[31][23]}{[12]} \right)^{h}
\left(\frac{[12]^2}{[23]^2} \right)^{h_1} \left(\frac{[12]^2}{[31]^2} \right)^{h_2}\\
&\qquad=\frac{[31][23]}{[12]} e^{\ell \frac{[31][23]}{[12]}}\delta \left(\frac{[12]^2}{[23]^2}-1 \right)
\delta \left(\frac{[12]^2}{[31]^2}-1 \right)\\
&\qquad={[12]^3}e^{\ell [12]}\delta([12]-[23])\delta([12]-[31]).
\end{split}
\end{equation}
In the third line the delta functions have the support on $[12]=\pm [23]$, $[12]=\pm [31]$. The associated four contributions were combined into a single one in the fourth line by using the property (\ref{16may5}).
By reinstating the momentum-conserving delta function, we find 
\begin{equation}
\label{9may12}
A_3 = g{[12]^3}e^{\ell[12]}\delta([12]-[23])\delta([12]-[31])
\delta^4(\lambda_1\bar\lambda_1+
\lambda_2\bar\lambda_2+\lambda_3\bar\lambda_3).
\end{equation}

As for the two-point function, we will now make a couple of simplifications, by changing the arguments of delta functions. First, one can note that 
\begin{equation}
\label{16jul9}
 [12]-[23]=0, \qquad  [12]-[31]=0
\end{equation}
is equivalent to
\begin{equation}
\label{16jul10}
 \bar\lambda_1^{\dot 1}+\bar\lambda_2^{\dot 1}+\bar\lambda_3^{\dot 1}=0, \qquad  \bar\lambda_1^{\dot 2}+\bar\lambda_2^{\dot 2}+\bar\lambda_3^{\dot 2}=0.
\end{equation}
 By passing to the new arguments in delta functions -- see appendix \ref{app:jac} for details -- we obtain 
\begin{equation}
\label{16jul12}
A_3=g{[12]^2}e^{\ell[12]}\delta^2(\bar\lambda_1+\bar\lambda_2+\bar\lambda_3) \delta^4(\lambda_1\bar\lambda_1+ \lambda_2\bar\lambda_2+\lambda_3\bar\lambda_3).
\end{equation}

On the next step, we note that 
\begin{equation}
\label{16jul13}
\begin{split}
 \lambda^{1}_1\bar\lambda^{\dot 1}_1+ \lambda^{1}_2\bar\lambda^{\dot 1}_2+\lambda^1_3\bar\lambda^{\dot 1}_3=0,\\
 \lambda^{1}_1\bar\lambda^{\dot 2}_1+ \lambda^{1}_2\bar\lambda^{\dot 2}_2+\lambda^1_3\bar\lambda^{\dot 2}_3=0,\\
 \lambda^{2}_1\bar\lambda^{\dot 1}_1+ \lambda^{2}_2\bar\lambda^{\dot 1}_2+\lambda^2_3\bar\lambda^{\dot 1}_3=0,\\
 \lambda^{2}_1\bar\lambda^{\dot 2}_1+ \lambda^{2}_2\bar\lambda^{\dot 2}_2+\lambda^2_3\bar\lambda^{\dot 2}_3=0
\end{split}
\end{equation}
is equivalent to 
\begin{equation}
\label{16jul14}
\begin{split}
\lambda_1^1 - \lambda_3^1=0, \qquad  \lambda_1^2-\lambda_3^2=0,\\
\lambda_2^1 - \lambda_3^1=0, \qquad  \lambda_2^2-\lambda_3^2=0
\end{split}
\end{equation}
on the support of the first delta function in (\ref{16jul12}). By passing to the new variables -- see appendix \ref{app:jac} for the relevant Jacobians -- we obtain
\begin{equation}
\label{16jul16}
A_3=ge^{\ell [12]}\delta^2(\bar\lambda_1+\bar\lambda_2+\bar\lambda_3) \delta^2(\lambda_2-\lambda_3)\delta^2(\lambda_1-\lambda_3).
\end{equation}
This is the form of the total three-point amplitude we were looking for. In the following, for brevity, we will drop the overall coupling constant $g$.

\section{Algebraic structures and higher-point amplitudes}
\label{sec:5}

The higher-spin symmetry of amplitudes in the AdS space can be made manifest, once these are presented as invariant traces of the higher-spin algebra. Below, we will show that both $A_2$ and $A_3$, that we computed above -- (\ref{9may10}) and (\ref{16jul16}) --  can be presented in the form of  invariant traces of 
a certain  flat space higher-spin algebra. Not only this makes the higher-spin symmetry of these amplitudes manifest, it also allows us to construct manifestly higher-spin symmetric higher-point amplitudes. 

To start, we introduce the inner product, associated with the two-point amplitude (\ref{9may10})
\begin{equation}
\label{6may2}
G_2\equiv (\Phi_1,\Phi_2) \equiv
\int d^2\lambda_1d^2\bar\lambda_1d^2\lambda_2d^2\bar\lambda_2 \delta^2(\lambda_1-\lambda_2)\delta^2(\bar\lambda_1+\bar\lambda_2)
\Phi_1(\lambda_1,\bar\lambda_1)
\Phi_2(\lambda_2,\bar\lambda_2).
\end{equation}
Next, we introduce the following binary product 
\begin{equation}
\label{6may1}
\begin{split}
(\Phi_1\ltimes\Phi_2)(\lambda_3,\bar\lambda_3)
&\equiv\int d^2\lambda_1d^2\bar\lambda_1d^2\lambda_2d^2\bar\lambda_2 \Phi_1(\lambda_1,\bar\lambda_1)
\Phi_2(\lambda_2,\bar\lambda_2)
\\
&\qquad \qquad e^{\ell [12]}\delta^2(\bar\lambda_1+\bar\lambda_2-\bar\lambda_3) \delta^2(\lambda_2-\lambda_3)\delta^2(\lambda_1-\lambda_3).
\end{split}
\end{equation}
As it is not hard to see, 
\begin{equation}
\label{6may3}
\begin{split}
G_3 =(\Phi_3,\Phi_1\ltimes\Phi_2)= \int d^2\lambda_1d^2\bar\lambda_1d^2\lambda_2d^2\bar\lambda_2 d^2\lambda_3d^2\bar\lambda_3
\Phi_1(\lambda_1,\bar\lambda_1)
\Phi_2(\lambda_2,\bar\lambda_2)
\Phi_3(\lambda_3,\bar\lambda_3)
 \\
e^{\ell [12]}\delta^2(\bar\lambda_1+\bar\lambda_2+\bar\lambda_3) \delta^2(\lambda_2-\lambda_3)\delta^2(\lambda_1-\lambda_3),
\end{split}
\end{equation}
that is the integral kernel of (\ref{6may3}) is the tree-point amplitude (\ref{16jul16}).

We note that the inner product (\ref{6may2}) can be represented as
\begin{equation}
\label{6may4}
(\Phi_1,\Phi_2)={\rm tr_{\ltimes}}(\Phi_1\ltimes\Phi_2),
\end{equation}
where 
\begin{equation}
\label{25oct8}
{\rm tr}_{\ltimes}(\Phi(\lambda,\bar\lambda))\equiv\int d^2\lambda d^2 \bar\lambda \Phi(\lambda,\bar\lambda)\delta^2(\bar\lambda).
\end{equation}
This leads to
\begin{equation}
\label{25oct9}
G_2 = {\rm tr_{\ltimes}}(\Phi_1\ltimes\Phi_2), \qquad G_3={\rm tr_{\ltimes}}(\Phi_1\ltimes(\Phi_2\ltimes\Phi_3)).
\end{equation}

It is simple to see that the $\ltimes$-product is associative. Indeed, on $\lambda$ variables it is just the usual commutative multiplication, while on $\bar\lambda$ variables it is related to the $\star$-product, employed in the definition of the AdS${}_4$ higher-spin algebra, by the Fourier transform (\ref{18may4}). 
Because of that, in (\ref{25oct9}) we are allowed to drop brackets indicating the order of multiplication.
 Besides that, it is straightforward to see from the explicit formula (\ref{6may2}), that the trace (\ref{6may4}) is cyclic for bosonic fields
\begin{equation}
\label{11may1}
{\rm tr_{\ltimes}}(\Phi_1\ltimes\Phi_2)={\rm tr_{\ltimes}}(\Phi_2\ltimes\Phi_1).
\end{equation}
Together with the associativity of the $\ltimes$-product, this implies that (\ref{25oct9}) are invariant with respect to the Lie algebra defined by the commutator
\begin{equation}
\label{11may2}
\bar\delta_{\varepsilon}\Phi \equiv  [\Phi,\varepsilon]_{\ltimes} \equiv  \Phi \ltimes \varepsilon-\varepsilon  \ltimes \Phi.
\end{equation}

We would like to remark that relevance of the product (\ref{6may1}) in the context of chiral higher-spin theories was already highlighted before. To be more precise, in \cite{Ponomarev:2017nrr} the off-shell version of (\ref{6may1}) was identified as the generalised ''colour algebra'' for chiral higher-spin theories in the sense of colour-kinematics duality and the BCJ relations \cite{Bern:2008qj}  applied to the self-dual sector \cite{Monteiro:2011pc}. Moreover, once the chiral higher-spin theory is rewritten as the self-dual Yang-Mills theory in the Chalmers-Siegel form \cite{Chalmers:1996rq}, the off-shell version of (\ref{6may1}) plays the role of the self-dual Yang-Mills gauge algebra. Algebra (\ref{6may1}) also features the twistor reformulation of the chiral higher-spin theory \cite{Krasnov:2021nsq,Krasnov:2021cva} as well as  recent works on its  free differential algebra reformulation \cite{Skvortsov:2022syz,Sharapov:2022dfw}.

In a similar manner one can construct invariants of  algebra (\ref{11may2}) with higher numbers of fields. Namely, for general $n$ 
\begin{equation}
\label{11may3}
G_n \equiv {\rm tr_{\ltimes}}(\Phi_1\ltimes\Phi_2 \ltimes \dots \ltimes \Phi_n)
\end{equation}
is invariant under (\ref{11may3}). Based on invariance with respect to higher-spin algebra symmetries, (\ref{11may3}) can be considered as a candidate higher-spin theory amplitude  in flat space. Before that, however, one may need to Bose symmetrise (\ref{11may3}) as, by construction, these invariants have only the cyclic symmetry. 

These results can be rewritten in a slightly different form, which is more reminiscent to the form of the known results in the AdS space. To this end, we pass to the Fourier transformed wave functions in the $\lambda$ variable 
\begin{equation}
\label{7may3}
\Phi(\lambda,\bar\lambda) =\frac{1}{4\pi^2} \int d^2\mu e^{i\mu\lambda}  \Psi(\mu,\bar\lambda), \qquad  \Psi(\mu,\bar\lambda) \equiv
\int d^2\lambda e^{i\lambda\mu}  \Phi(\lambda,\bar\lambda).
\end{equation}
One then finds that 
\begin{equation}
\label{6may10}
{\rm tr}_{\ltimes}(\Phi)= {\rm tr}(\Psi),
\end{equation}
where 
\begin{equation}
\label{6may10x1}
{\rm tr}(\Psi)\equiv \int d^2\lambda d^2 \bar\lambda \Psi(\lambda,\bar\lambda)\delta^2(\bar\lambda)\delta^2(\lambda).
\end{equation}
To arrive to (\ref{6may10}), we used 
\begin{equation}
\label{6may22}
\delta^2(\lambda) = \frac{1}{4\pi^2}\int d^2\mu e^{i\mu \lambda} .
\end{equation}
Note that unlike the previous trace, ${\rm tr}$ is parity-invariant. It is also the trace that is used in the AdS space case, see below.

The $\ltimes$-product for fields $\Phi$ induces another associative product on the Fourier transformed wave functions
\begin{equation}
\label{11may4}
\Psi_1 \triangleright \Psi_2 \equiv \Psi_3, \qquad \text{where} \qquad \Phi_3 = \Phi_1\ltimes \Phi_2.
\end{equation}
From the explicit computation, we find
\begin{equation}
\label{7may9}
\begin{split}
 ( \Psi_1 \triangleright\Psi_2)(\mu_3,\bar\lambda_3)
&=\frac{1}{4\pi^2}\int d^2\mu_1d^2\bar\lambda_1d^2\mu_2d^2\bar\lambda_2
\Psi_1(\mu_1,\bar\lambda_1)
 \Psi_2(\mu_2,\bar\lambda_2)
\\
&\qquad\qquad\qquad\qquad e^{\ell [12]}\delta^2(\bar\lambda_1+\bar\lambda_2-\bar\lambda_3) \delta^2(\mu_1+\mu_2-\mu_3).
\end{split}
\end{equation}
In these terms, invariants (\ref{11may3}) can be presented as 
\begin{equation}
\label{11may5}
G_n \equiv {\rm tr}( \Psi_1\triangleright \Psi_2 \triangleright \dots \triangleright  \Psi_n).
\end{equation}

Yet another useful representation can be obtained by making the Fourier transform in the $\bar\lambda$ variable 
\begin{equation}
\label{18may2}
\Phi(\lambda,\bar\lambda) =\frac{1}{4\pi^2} \int d^2\bar\mu e^{i\bar\mu\bar\lambda}  \Upsilon(\lambda,\bar\mu), \qquad  \Upsilon(\lambda,\bar\mu) \equiv
\int d^2\bar\lambda e^{i\bar\lambda\bar\mu}  \Phi(\lambda,\bar\lambda).
\end{equation}
In these terms, the trace becomes
\begin{equation}
\label{18may3}
{\rm tr}_{\ltimes}(\Phi)= {\rm tr}_{\bar\circ}(\Upsilon)\equiv \frac{1}{4\pi^2} \int d^2\lambda d^2\bar\mu \Upsilon(\lambda,\bar\mu)
\end{equation}
and the associative product reads
\begin{equation}
\label{18may4}
\begin{split}
 ( \Upsilon_1 \bar\circ\Upsilon_2)(\lambda_3,\bar\mu_3)
&\equiv\frac{1}{4\pi^2\ell}\int d^2\lambda_1d^2\bar\mu_1d^2\lambda_2d^2\bar\mu_2
\Upsilon_1(\lambda_1,\bar\mu_1)
 \Upsilon_2(\lambda_2,\bar\mu_2)
\\
&\qquad\qquad\qquad\qquad e^{\frac{1}{\ell} ([\mu_1\mu_2]+[\mu_2\mu_3]+[\mu_3\mu_1])} \delta^2(\lambda_2-\lambda_3)\delta^2(\lambda_1-\lambda_3).
\end{split}
\end{equation}
This representation is more suitable for studying the higher-spin algebra, as in these terms the Poincare algebra is generated by polynomial parameters, see below.

\section{Higher-spin invariants in the AdS space}
\label{sec:6}

In this section we briefly review the construction of  higher-spin invariant amplitudes in the AdS space \cite{Colombo:2012jx,Didenko:2012tv,Gelfond:2013xt,Didenko:2013bj} with the main goal being to highlight similarity of their construction and the presentation given above for the flat space amplitudes. 

The higher-spin algebra in AdS${}_4$ is defined by the star product commutator\footnote{Many ingredients used here are standard in the higher-spin literature, see e.g. \cite{Vasiliev:1999ba} for review.} 
\begin{equation}
\label{11may6}
[\Upsilon_1,\Upsilon_2]_{\star} = \Upsilon_1\star \Upsilon_2 - \Upsilon_2\star \Upsilon_1,
\end{equation}
where the $\star$ product is given by
\begin{equation}
\label{7may10}
\begin{split}
 (\Upsilon_1 \star\Upsilon_2)(y_3,\bar y_3)
\equiv&\int d^2 y_1d^2\bar y_1d^2 y_2d^2\bar y_2
\Upsilon_1(y_1,\bar y_1)
\Upsilon_2( y_2,\bar y_2)\\
&\qquad \qquad\qquad \qquad
e^{i([y_2y_1]+[y_1y_3]+[y_3y_2])}
e^{i(\langle y_2y_1\rangle+\langle y_1y_3\rangle+\langle y_3y_2\rangle)}
.
\end{split}
\end{equation}
Here we use the standard notation $y$ for the spinor variables. These will be connected to $\lambda$'s and $\mu$'s from the previous section below.

For bosonic fields this associative product admits an invariant trace (\ref{6may10x1}), which means that 
\begin{equation}
\label{11may6x1}
G_n\equiv {\rm tr}(\Upsilon_1 \star \Upsilon_2\star \dots \star \Upsilon_n)
\end{equation}
is invariant with respect to higher-spin symmetries
\begin{equation}
\label{11may6x2}
\delta_\xi \Upsilon = [\Upsilon,\xi]_{\star}.
\end{equation} 

Massless fields transform in the so-called twisted adjoint representation of the higher-spin algebra, which is defined by
\begin{equation}
\label{11may7}
\delta_\xi \Omega = - \xi \star \Omega + \Omega \star \tilde\xi,
\end{equation}
where 
\begin{equation}
\label{11may8}
\tilde \xi(y,\bar y)\equiv \xi(- y,\bar y)=\xi ( y,-\bar y)
\end{equation}
and the last equality is true for bosonic fields. 

It is straightforward to check that the star product with the delta function 
\begin{equation}
\label{11may9}
\Upsilon = \Omega \star \delta^2( \bar y)
\end{equation}
performs the Fourier transform in $\bar y$ variable. 
The following identities can be also easily derived
\begin{equation}
\label{19may2}
\Omega \star \delta^2(y)\star \delta^2( y)\equiv \Omega, \qquad \Omega \star \delta^2(\bar y)\star \delta^2(\bar y)\equiv\Omega.
\end{equation}
Moreover, the twist in (\ref{11may8}) can be implemented via
\begin{equation}
\label{19may1}
\xi(y, -\bar y)\equiv\delta^2( \bar y)\star \xi( y,\bar y)\star \delta^2( \bar y).
\end{equation}

These identities allow one to show that once $\Omega$ transforms in the twisted adjoint representation (\ref{11may7}), then $\Upsilon$ in (\ref{11may9}) transforms in the adjoint representation (\ref{11may6x2}), \cite{Didenko:2009td}. Accordingly, trace
\begin{equation}
\label{11may10}
G_n \equiv {\rm tr}( \Omega_1 \star \delta^2(\bar y) \star  \Omega_2 \star \delta^2(\bar y)\star \dots \star \Omega_n \star \delta^2(\bar y))
\end{equation}
is  invariant with $\Omega$'s transforming in the twisted adjoint representation of the AdS higher-spin algebra. For example, the AdS invariant (\ref{11may10}) in the three-point case  gives 
\begin{equation}
\label{18may5}
\begin{split}
G_3 = &\int d^2 y_1d^2\bar y_1d^2 y_2d^2\bar y_2 d^2 y_3d^2\bar y_3
\Omega_1(y_1,\bar y_1)
\Omega_2(y_2,\bar y_2)
\Omega_3(y_3,\bar y_3)
 \\
& \qquad\qquad\qquad\qquad\qquad\qquad\qquad e^{i([ y_2y_1]+[ y_1y_3]+[ y_3y_2])} e^{i \langle y_1 y_2\rangle }\delta^2( y_1+ y_2+ y_3) 
\end{split}
\end{equation}
and its kernel
\begin{equation}
\label{4aug1}
A_3\equiv e^{i([ y_2y_1]+[ y_1y_3]+[ y_3y_2])} e^{i \langle y_1 y_2\rangle }\delta^2( y_1+ y_2+ y_3) 
\end{equation}
can be identified as the AdS space three-point amplitude.

It is now easy to see similarities between (\ref{11may5}) and (\ref{11may6x1}), as well as between (\ref{11may3}) and (\ref{11may10}).
Indeed, to obtain the flat space result from (\ref{11may6x1}) one just needs to replace the $\star$-product with the $\triangleright$-product. Similarly, the counterpart of the $\ltimes$-product in (\ref{11may3}) is the composition $\star \delta^2(\bar y) \star$, while 
${\rm tr}(\Omega \star \delta^2(\bar y))$ corresponds to $ {\rm tr}_{\ltimes}(\Phi)$.
Similar comparison can be made for explicit expressions for the amplitudes. 

In fact, the flat space and the AdS space results may have not only a structural similarity, but also they may be related by a contraction. Let us illustrate this with the example of the three-point amplitude (\ref{4aug1}). To start, we recall, that the arguments $(y,\bar y)$ and $(\lambda,\bar\lambda)$ of the higher-spin fields are related via $y\sim \frac{\partial}{\partial \lambda}$, $\bar y\sim \frac{\partial}{\partial \bar \lambda}$. Accordingly, $\Omega(y,\bar y)$ is the double Fourier transform of $\Phi(\lambda,\bar\lambda)$ on  $\lambda$ and $\bar\lambda$. Performing such a double Fourier transform with (\ref{4aug1}), we find 
\begin{equation}
\label{4aug2}
A_3 = e^{i [ 1 2 ] }\delta^2(\bar \lambda_1+\bar \lambda_2+\bar \lambda_3) e^{i(\langle 21\rangle+\langle 13\rangle+\langle 32\rangle)}.
\end{equation}
Next, we introduce a pair of independent AdS radii -- $R$ and $\bar R$ -- as
\begin{equation}
\label{4aug3}
A_3(R,\bar R) \equiv  e^{\frac{i}{\bar R} [ 1 2 ] }\delta^2(\bar \lambda_1+\bar \lambda_2+\bar \lambda_3) e^{iR(\langle 21\rangle+\langle 13\rangle+\langle 32\rangle)}.
\end{equation}
Normally, in the AdS space case one has $R=\bar R$, while we consider to different radii, as we are going to make a chiral contraction $R\to \infty$. 
Since, (\ref{4aug3}) is not analytic in the cosmological constant $\Lambda \sim \frac{1}R$, it may seem that its limit $\Lambda \to 0$ is
 ill-defined. This parallels the standard argument that the flat space limit of higher-spin theories does not exist, see e.g. \cite{Fradkin:1987ks}\footnote{ In \cite{Boulanger:2008tg} it was shown that the flat space limit can be made smooth for individual cubic vertices. Still, up to now it was not clear how to take the flat space limit for the complete cubic action.}. Note, however, that the $R$ dependence of (\ref{4aug3}) is just the dependence of the star product on the non-commutativity parameter. In the limit $R\to \infty$ the star product becomes commutative, which can also be seen by making the Fourier transform in $\lambda$ before evaluating the limit and then Fourier transforming the result back. Explicitly, one has
\begin{equation}
\label{4aug4}
\lim_{R\to \infty} R e^{iR(\langle 21\rangle+\langle 13\rangle+\langle 32\rangle)} =  \delta^2(\lambda_2-\lambda_3)\delta^2(\lambda_1-\lambda_3).
\end{equation}
Identifying, in addition, $\frac{i}{\bar R}=\ell$, we find that the flat space limit of (\ref{4aug3}), indeed, reproduces the flat result (\ref{16jul16}). It would be interesting to see whether this limiting procedure can be extended to higher-point invariants. If so, it can be used to obtain the relative overall factors of the flat space amplitudes from those in AdS by using  the flat space limit.

\section{Properties}
\label{sec:7}

In the present section we discuss simple properties of amplitudes that we defined in (\ref{11may3}).

To start, we evaluate explicitly (\ref{11may3}). It is more convenient to do that in the form 
\begin{equation}
\label{16may16}
\begin{split}
G_n &= (\Phi_1,\Phi_2 \ltimes \dots \ltimes \Phi_n) \\
&\qquad = 
\int \prod_{i=1}^n d^2\lambda_i d^2\bar\lambda_i \Phi_i(\lambda_i,\bar\lambda_i)
\prod_{n\ge i>j\ge 2} e^{\ell [ji]}
\delta^2(\sum_{i=1}^n \bar\lambda_i)\prod_{i=2}^{n}\delta^2(\lambda_1 -\lambda_i).
\end{split}
\end{equation}
In particular, for four fields one finds
\begin{equation}
\label{16may17}
A_4 = e^{\ell([23]+[24]+[34])}\delta^2(\bar\lambda_1+\bar\lambda_2+\bar\lambda_3+\bar\lambda_4)\delta^2(\lambda_1-\lambda_2)\delta^2(\lambda_1-\lambda_3)\delta^2(\lambda_1-\lambda_4).
\end{equation}
These amplitudes have quite an unusual structure: the scattering occurs only for momenta with the same $\lambda$ components, while momentum conservation is achieved due to the fact that $\bar\lambda$ is conserved separately. In terms of momenta -- which should necessarily be complex for (\ref{16may16}) and (\ref{16may17}) to be non-trivial -- this entails that only fields with collinear momenta scatter. In terms of the Mandelstam variables, this implies that the amplitude is distributional with all independent Mandelstam variables set to zero. This type of amplitudes has already occurred in the higher-spin literature for conformal higher-spin theories as well as for  Mellin amplitudes in holographic higher-spin theories. It was also anticipated that this type of amplitudes may occur for massless higher-spin theories in flat space and above we provided a concrete example to support these expectations. 

By construction, amplitudes (\ref{11may3}) are invariant with respect to (\ref{11may2}). It is demonstrated in appendix \ref{app:poin}  that parameters
\begin{equation}
\label{16may18}
\varepsilon_1 \sim \lambda_\alpha \frac{\partial}{\partial\bar\lambda^{\dot\alpha}}  \delta^2(\bar\lambda) \qquad \text{and} \qquad \varepsilon_2\sim 
\left(\frac{\partial}{\partial\bar\lambda^{\dot\alpha}} \right)^2 \delta^2(\bar\lambda)
\end{equation}
generate $P$ and $\bar J$ of the Poincare algebra (\ref{16may2}). At the same time, invariance of (\ref{11may3}) under Lorentz transformations of opposite chirality $J$, though, being manifest, does not follow from invariance under (\ref{11may2}).

It is convenient to consider the symmetry parameters in the Fourier transformed representation with respect to  $\bar\lambda$
\begin{equation}
\label{16may19}
\varepsilon(\lambda,\bar\lambda) =\frac{1}{4\pi^2} \int d^2\bar\mu e^{i\bar\mu\bar\lambda}  \xi(\bar\mu,\lambda), \qquad  \xi(\bar\mu,\lambda) \equiv
\int d^2\bar\lambda e^{i\bar\lambda\bar\mu}  \varepsilon(\lambda,\bar\lambda).
\end{equation}
In this representation Poincare generators $P$ and $\bar J$ are represented by quadratic expressions
\begin{equation}
\label{16may20}
 \xi_1\sim \lambda_\alpha \bar\mu_{\dot\alpha}  \qquad \text{and} \qquad \xi_2 \sim \bar\mu_{\dot\alpha}\bar\mu_{\dot\alpha}.
\end{equation}
The $\ltimes$-product upon the Fourier transform in $\bar\lambda$ goes into the star product in $\bar\mu$ and the trivial product in $\lambda$, (\ref{18may4}). Accordingly, $\xi_1$ and $\xi_2$ on general $\xi$ generate the following action of the Poincare generators
\begin{equation}
\label{16may21}
P_{\alpha\dot\alpha} \xi(\lambda,\bar\mu) \sim \lambda_{\alpha}\frac{\partial}{\partial \bar\mu^{\dot\alpha}}\xi(\lambda,\bar\mu), \qquad 
\bar J_{\dot\alpha\dot\alpha}\xi(\lambda,\bar\mu)\sim \bar\mu_{\dot\alpha}\frac{\partial}{\partial \bar\mu^{\dot\alpha}}\xi(\lambda,\bar\mu).
\end{equation}
One can also show that $J$ transformations of $\Phi$ via
\begin{equation}
\label{16may22}
\bar\delta_{\varepsilon(\xi)}\Phi = [\Phi,\varepsilon(\xi)]_{\ltimes},
\end{equation}
where $\varepsilon(\xi)$ refers to the Fourier transform (\ref{16may19}), induce $J$ transformations on $\xi$
\begin{equation}
\label{16may23}
 J_{\alpha\alpha}\xi(\lambda,\bar\mu)\sim \lambda_{\alpha}\frac{\partial}{\partial\lambda^{\alpha}}\xi(\lambda,\bar\mu).
\end{equation}
Formulas (\ref{16may21}), (\ref{16may23}) define how the higher-spin algebra parameters $\xi$ transform under the Poincare algebra.

In the covariant formalism one can show that parameters of the global higher-spin algebra should transform as traceless Killing tensors under the global isometry algebra, see e.g. \cite{Vasiliev:1986td,Bekaert:2005ka}.
In the 4d Minkowski space this is consistent with (\ref{16may21}), (\ref{16may23}) for $\xi$ with homogeneity degrees $\bar N > N$, while for $\bar N= N$ translations should act trivially and for $\bar N < N$ these should act by  $P \sim \lambda \partial_{\bar\mu}$ to reproduce transformations of the traceless Killing tensors. 
In other words, $\xi$ does not transform as normally expected in covariant theories. 
Nevertheless, in recent works \cite{Krasnov:2021nsq,Krasnov:2021cva,Skvortsov:2022syz,Sharapov:2022dfw}, where the chiral higher-spin theory was covariantised, (\ref{16may21}), (\ref{16may23}) was still achieved at the expense of loosing parity invariance. 

Together with the quadratic parameters (\ref{16may20}), one can also consider 
\begin{equation}
\label{5aug1}
\xi_3 \sim \lambda_\alpha \lambda_{\alpha}.
\end{equation}
Let the associated symmetry generator be $L$. Then, as it is not hard to see, $\xi_1$, $\xi_2$ and $\xi_3$ generate an algebra of the following schematic form
\begin{equation}
\label{5aug2}
\begin{split}
[\bar J,\bar J] &\sim \bar J, \qquad [\bar J, P]\sim P, \qquad [P,P]\sim L,\\
[L,\bar J] &=0, \qquad [L,P]=0, \qquad \; [L,L]=0.
\end{split}
\end{equation}
Note that $[P,P]\ne 0$.
This algebra can be regarded as a central extension of the chiral part of the Poincare algebra, the latter generated by $P$ and $\bar J$. Alternatively, (\ref{5aug2}) can be obtained as the following chiral contraction of $so(3,2)$ 
\begin{equation}
\label{5aug3}
\bar J \to \bar J, \qquad P \to \frac{1}{R} P, \qquad J \to \frac{1}{R^2}L \qquad \text{where} \qquad R\to \infty.
\end{equation}

As mentioned above, though, $J$ is not part of symmetry (\ref{11may2}), it still is  a symmetry of amplitudes. It acts on generators $P$, $\bar J$ and $L$ according to the index structure of the latter
\begin{equation}
\label{5aug4}
[J,P]\sim P, \qquad [J,\bar J]=0, \qquad  [J,L]\sim L, \qquad   [J,J] \sim J.
\end{equation}
Note that  (\ref{5aug2}) together with (\ref{5aug4}) is not  a central extension of the complete Poincare algebra, since $L$ commutes with $J$ non-trivially\footnote{As is well-known, the Poincare algebra does not admit non-trivial central extensions \cite{Weinberg}.}. It is also worth remarking that, due to the fact that $\lambda$'s commute, $L$ acts trivially on the higher-spin fields, $L=0$. This means that when dealing with massless higher-spin fields (\ref{5aug2}), (\ref{5aug4}), in effect, reduces to the usual Poincare algebra.

Finally, let us comment on parity of the AdS space and of the flat space higher-spin amplitudes. 
As we explained, the AdS amplitude
(\ref{11may10}) is manifestly invariant under (\ref{11may7}), 
\begin{equation}
\label{19may3}
\delta_\xi \Phi = - \xi \star \Phi + \Phi \star \delta^2(\bar y)\star \xi \star \delta^2(\bar y).
\end{equation}
For bosonic fields one has (\ref{11may8}), so this symmetry transformation can also be written as
\begin{equation}
\label{19may4}
\delta_\xi \Phi = - \xi \star \Phi + \Phi \star \delta^2( y)\star \xi \star \delta^2( y).
\end{equation}
By the very same logic (\ref{19may3}), (\ref{19may4}) leaves 
\begin{equation}
\label{17may1}
\bar G_n \equiv {\rm tr}( \Phi_1 \star \delta^2(y) \star  \Phi_2 \star \delta^2(y)\star \dots \star \ \Phi_n \star \delta^2(y))
\end{equation}
invariant, which means that by adding (\ref{11may10}) and (\ref{17may1}), one can restore parity without violating higher-spin symmetry. 

Let us now move to the flat space case. Along with (\ref{11may3}), one can consider its complex conjugate
\begin{equation}
\label{17may6}
G_n \equiv {\rm tr_{\rtimes}}(\Phi_1\rtimes\Phi_2 \rtimes \dots \rtimes \Phi_n),
\end{equation}
which is defined by means of the complex conjugate associative product 
\begin{equation}
\label{17may7}
\begin{split}
(\Phi_1\rtimes\Phi_2)(\lambda_3,\bar\lambda_3)
&\equiv\int d^2\lambda_1d^2\bar\lambda_1d^2\lambda_2d^2\bar\lambda_2
\Phi_1(\lambda_1,\bar\lambda_1)
\Phi_2(\lambda_2,\bar\lambda_2)
\\
&\qquad\qquad\qquad e^{\ell\langle 12\rangle}\delta^2(\lambda_1+\lambda_2-\lambda_3) \delta^2(\bar\lambda_2-\bar\lambda_3)\delta^2(\bar\lambda_1-\bar\lambda_3)
.
\end{split}
\end{equation}
Amplitudes (\ref{17may6}) are manifestly invariant with respect to
\begin{equation}
\label{17may9}
\delta_{\varepsilon}\Phi = [\Phi,\varepsilon]_{\rtimes} \equiv  \Phi \rtimes \varepsilon-\varepsilon  \rtimes \Phi.
\end{equation}

We then make
a straightforward  check that  variation of the three-point amplitude (\ref{16jul16}) under 
(\ref{17may9})
gives
\begin{equation}
\begin{split}
\label{a17may1}
\delta_\varepsilon A_3 =&\int d^2\lambda d^2\bar\lambda \varepsilon(\lambda,\bar\lambda)
\\
 \Big(&e^{\ell [12]}\delta^2(\bar\lambda_1+\bar\lambda_2+\bar\lambda_3)\delta^2 (\lambda_2-\lambda_3)\delta^2(\lambda_1+\lambda - \lambda_3)
(e^{\ell\langle 1\lambda\rangle}- e^{\ell\langle \lambda 1\rangle})\delta^2(\bar\lambda-\bar\lambda_1)\\
&+e^{\ell[12]}\delta^2(\bar\lambda_1+\bar\lambda_2+\bar\lambda_3)\delta^2 (\lambda_1-\lambda_3)\delta^2(\lambda_2+\lambda - \lambda_3)
(e^{\ell\langle 2\lambda\rangle}- e^{\ell\langle \lambda 2\rangle})\delta^2(\bar\lambda-\bar\lambda_2)\\
&+e^{\ell[12]}\delta^2(\bar\lambda_1+\bar\lambda_2+\bar\lambda_3)\delta^2 (\lambda_1-\lambda_2)\delta^2( \lambda_3+\lambda -\lambda_2)
(e^{\ell\langle 3\lambda\rangle}- e^{\ell\langle \lambda 3\rangle})\delta^2(\bar\lambda-\bar\lambda_3)\Big),
\end{split}
\end{equation}
which is non-vanishing for general $\varepsilon$. Indeed, three terms in (\ref{a17may1}) are supported on three different channels 
\begin{equation}
\label{17may10}
\begin{split}
1: \qquad \lambda_2=\lambda_3, \qquad \bar\lambda=\bar\lambda_1,\\
2: \qquad \lambda_1=\lambda_3, \qquad \bar\lambda=\bar\lambda_2,\\
3: \qquad \lambda_1=\lambda_2, \qquad \bar\lambda=\bar\lambda_3,
\end{split}
\end{equation}
so they do not cancel out for general $\varepsilon$. 
Moreover, this variation cannot cancel against the variation of the parity conjugate amplitude under (\ref{11may2}) with any parameter $\varepsilon'(\varepsilon)$ as the latter variation is supported on yet another three channels
\begin{equation}
\label{19may6}
\begin{split}
4: \qquad \bar\lambda_2=\bar\lambda_3, \qquad \lambda=\lambda_1,\\
5: \qquad \bar\lambda_1=\bar\lambda_3, \qquad \lambda=\lambda_2,\\
6: \qquad \bar\lambda_1=\bar\lambda_2, \qquad \lambda=\lambda_3.
\end{split}
\end{equation}
We, therefore, conclude that  
\begin{equation}
\label{17may11}
{\rm tr_{\ltimes}}(\Phi_1\ltimes\Phi_2 \ltimes \Phi_3)+{\rm tr_{\rtimes}}(\Phi_1\rtimes\Phi_2  \rtimes \Phi_3)
\end{equation}
is not invariant with respect to (\ref{11may2}) and (\ref{17may9}), even if the transformation parameters of the two transformations are not independent, so such a naive parity-invariant completion does not work. 

It should be remarked, however, that in the same way as it was discussed around (\ref{16may18}), it can be shown that (\ref{17may9}) with
\begin{equation}
\label{17may12}
\varepsilon_1 \sim \bar\lambda_{\dot\alpha} \frac{\partial}{\partial\lambda^{\alpha}}  \delta^2(\lambda) \qquad \text{and} \qquad \varepsilon_2\sim 
\left(\frac{\partial}{\partial\lambda^{\alpha}} \right)^2 \delta^2(\lambda)
\end{equation}
generate Poincare transformations $P_{\alpha\dot\alpha}$ and $J_{\alpha\alpha}$. These leave invariant all amplitudes of both chiralities, so, in particular, with these $\varepsilon$  
(\ref{a17may1}) vanishes. In other words, despite amplitudes are not invariant with respect to general transformations of opposite chirality, for some parameters invariance does hold. It would be interesting to explore this issue in more detail. Finally, we make an obvious comment that 
\begin{equation}
\label{17may13}
{\rm tr_{\ltimes}}(\Phi_1\ltimes\Phi_2)={\rm tr_{\rtimes}}(\Phi_1\rtimes\Phi_2 ),
\end{equation}
which means that the two-point function is invariant with respect to transformations of both types.

\section{Conclusion}
\label{sec:8}

In the present paper we considered two- and three-point amplitudes in the chiral higher-spin theory in flat space. Using a certain regularisation, we summed them over helicities. Resulting expressions quite manifestly have the form of invariant traces of products of  on-shell higher-spin fields. The product involved in these expressions is the associative product of the chiral flat space higher-spin algebra, while the trace is cyclic with respect to this product. These properties make invariance of these two- and three-point amplitudes under chiral flat space higher-spin algebra transformations manifest. Moreover, employing these ingredients, we construct invariant traces for higher numbers of fields. Being invariant with respect to the aforementioned higher-spin algebra, they serve as natural candidates for higher-point amplitudes of higher-spin gauge fields. These give the first example of non-trivial higher-spin scattering in flat-space beyond three points. It should be emphasised, however, that similarly to the three-point amplitude in the chiral theory, higher-point amplitudes we constructed are defined for complex momenta.

This flat space construction  closely mimics \cite{Colombo:2012jx,Didenko:2012tv,Gelfond:2013xt}, in which amplitudes for higher-spin gauge fields in the AdS space were constructed from the requirement of invariance with respect to the AdS space higher-spin algebra. Despite the AdS space amplitudes produced in this approach are superficially  chiral, with one extra step these can be made parity-invariant and, eventually, they can be matched with the correlators on the CFT side. 
In the present paper we were not able to extend this last step to the flat space case.

Flat space higher-spin amplitudes that we constructed have a peculiar analytic structure. More precisely, these are supported on kinematic configurations with one of the momentum spinors being equal for all fields, while momentum is conserved due to conservation of the momentum spinor of opposite chirality. In the usual vector language this translates into the statement that the scattering is non-trivial only for collinear momenta and, thereby, leads to distributional amplitudes. Relevance of distributional amplitudes in the higher-spin context has been already highlighted before. In particular, 
distributional amplitudes were found by the direct computation of the tree-level four-point amplitude in the conformal higher-spin theory in \cite{Joung:2015eny,Beccaria:2016syk}. It was also shown that this result follows from the invariance with respect to the conformal higher-spin algebra. In \cite{Sleight:2016xqq} this conclusion was extended to flat space amplitudes in massless theories, by assuming that the Lorentz part of the AdS higher-spin algebra in the flat space limit remains intact. Moreover, Mellin amplitudes for higher-spin theories computed holographically, also have distributional nature \cite{Taronna:2016ats,Bekaert:2016ezc}. Thus, there is growing evidence that  distributional structure of amplitudes in higher-spin gauge theories is unavoidable. Our results give concrete examples of amplitudes, supporting these expectations. 

There are several open problems that we would like to address in future. First of all, it is not clear what is the theory that amplitudes obtained in the present paper correspond to. In particular, it would be interesting to construct the associated action and to find out whether it is local or not. It is known that the chiral theory cannot be completed to the parity-invariant one without violating locality -- see \cite{Metsaev:1991mt} and the explicit analysis reviewed in \cite{Ponomarev:2017nrr}. At the same time, to the best of our knowledge, completions of the chiral theory of other types -- e.g. with higher-order chiral vertices -- have not been studied in full detail, so locality of the theory with the amplitudes that we constructed is not ruled out. In principle, the associated light-cone action can be constructed directly from amplitudes following the approach of \cite{Ponomarev:2016cwi}. Still, it would be interesting to carry out this analysis more explicitly. 

The amplitudes that we constructed feature an undetermined overall factor for every number of external lines. As in the AdS space case, these prefactors cannot be fixed purely from symmetry considerations. In  the AdS space, these can be determined either by the direct comparison with the CFT correlators or, from more general considerations, such as the consistency of the CFT correlators with the OPE. In the flat space limit the latter constraints translate into factorisation properties of flat-space amplitudes, which can be connected to unitarity \cite{Penedones:2010ue,Fitzpatrick:2011dm,Ponomarev:2019ofr,Meltzer:2019nbs}. It would be interesting to see whether these ideas can, indeed, fix the relative coefficients of different amplitudes, especially, considering the distributional and chiral nature of amplitudes involved. Moreover, the flat space higher-spin algebra underlying our analysis may be helpful in guessing the holographically dual description of the flat-space higher-spin theory.

Finally, the most exciting future problem is to render the amplitudes we suggested here  parity invariant. If we are to keep symmetries as the main guiding principle, the key missing ingredient on this way is the suitable parity-invariant flat-space higher-spin algebra. In this regard, we would like to mention, that just the Killing tensors, transforming in the appropriate representations of the Poincare algebra can already be regarded as the flat-space higher-spin algebra. Some more non-trivial examples of flat-space higher-spin algebras were constructed recently as the result of contraction of the AdS higher-spin algebra \cite{Campoleoni:2021blr}, while the associated singleton representation in the 4d case was constructed in \cite{Ponomarev:2021xdq}. A somewhat unattractive feature of this latter algebra is that parameters of higher-spin symmetries do not transform as Killing tensors under the Poincare algebra. In any case, it would be important to analyse higher-spin algebras in flat space more systematically, irrespectively of the contraction from the AdS space.

\acknowledgments

We would like to thank E. Skvortsov and V. Didenko for fruitful discussions on various subjects related to the paper and for comments on the draft. This work was supported by Russian Science Foundation Grant 18-12-00507.

\appendix

\section{Jacobians}
\label{app:jac}
In the appendix we collect the results for Jacobians, which were used in the main text for changing the arguments of delta functions. 

When showing that (\ref{28apr12}) is equivalent to (\ref{9may3}), we needed to go from $\delta(x_1)\delta(x_2)\delta(x_3)$ to $\delta(y_1)\delta(y_2)\delta(y_3)$
where
\begin{equation}
\label{28apr14}
\begin{split}
x_1& \equiv p_1^1+p_1^2 = -\frac{1}{2}(\bar\lambda_{\dot 1}^1 \lambda_2^1 +\bar\lambda_{\dot 2}^1 \lambda_1^1 +
\bar\lambda_{\dot 1}^2 \lambda_2^2 +\bar\lambda_{\dot 2}^2 \lambda_1^2),\\
x_2&\equiv p_2^1+p_2^2 = \frac{i}{2}(\bar\lambda_{\dot 1}^1 \lambda_2^1 -\bar\lambda_{\dot 2}^1 \lambda_1^1 +
\bar\lambda_{\dot 1}^2 \lambda_2^2 -\bar\lambda_{\dot 2}^2 \lambda_1^2),\\
x_3&\equiv p_3^1+p_3^2 = \frac{1}{2}(-\bar\lambda_{\dot 1}^1 \lambda_1^1 +\bar\lambda_{\dot 2}^1 \lambda_2^1 -
\bar\lambda_{\dot 1}^2 \lambda_1^2 +\bar\lambda_{\dot 2}^2 \lambda_2^2)
\end{split}
\end{equation}
and
\begin{equation}
\label{28apr7}
\begin{split}
y_1&\equiv\langle 12 \rangle =  \lambda^1_{1} \lambda^2_{2}-\lambda^1_{2} \lambda^2_{1},
\\
y_2&\equiv[ 12 ] =  \bar\lambda^1_{\dot 1} \bar\lambda^2_{\dot 2}-\bar\lambda^1_{\dot 2} \bar\lambda^2_{\dot 1},
\\
y_3&\equiv\langle 1\mu\rangle [\mu 1]+\langle 2\mu \rangle [\mu 2]= -\lambda^1_{1}\bar\lambda^1_{\dot 1}- \lambda^2_{1}\bar\lambda^2_{\dot 1},
\end{split}
\end{equation}
where we chose $\mu_1=0$, $\mu_2 =1$ and similarly for $\bar\mu$.
One can show that 
\begin{equation}
\label{28apr15x1}
\begin{split}
x_1 &= -\frac{1}{2} \left(-\frac{\lambda_2^1}{\lambda_1^1}y_3 + y_1\frac{\bar\lambda_{\dot 1}^2}{\lambda_1^1}-\frac{\bar\lambda_{\dot 2}^1}{\bar\lambda_{\dot 1}^1}y_3 
+ \frac{\lambda_1^2}{\bar\lambda_{\dot 1}^1 }y_2\right),\\
x_2 &= \frac{i}{2} \left(-\frac{\lambda_2^1}{\lambda_1^1}y_3 + y_1\frac{\bar\lambda_{\dot 1}^2}{\lambda_1^1}+\frac{\bar\lambda_{\dot 2}^1}{\bar\lambda_{\dot 1}^1}y_3 
- \frac{\lambda_1^2}{\bar\lambda_{\dot 1}^1 }y_2\right),\\
x_3& = \frac{1}{2}\left( 
-\frac{\lambda_2^1\bar\lambda_{\dot 2}^1}{\lambda_1^1 \bar\lambda_{\dot 1}^1}y_3 +y_3
+
\frac{y_1y_2}{\lambda_1^1 \bar\lambda_{\dot 1}^1}+
y_1 \frac{\bar\lambda_{\dot 2}^1\bar\lambda_{\dot 1}^2}{\lambda_1^1 \bar\lambda_{\dot 1}^1}
+ y_2 \frac{\lambda_2^1 \lambda_1^2}{\lambda_1^1 \bar\lambda_{\dot 1}^1}
\right).
\end{split}
\end{equation}
We need to evaluate 
\begin{equation}
\label{28apr15}
J_1\equiv \left|\frac{\partial (y_1,y_2,y_3)}{\partial (x_1,x_2,x_3)} \right|_{x_i=0}
\end{equation}
which has to be evaluated at  $x_i=0$, which is equivalent to $y_i=0$. To extract $y(x)$ from (\ref{28apr15x1}), we can drop the $y_1y_2$ in the last line, as we are interested only in the first derivatives of $y$'s with respect to $x$'s at $x_i=0$. Expressing $y$'s from (\ref{28apr15x1}) with the quadratic terms dropped and evaluating the Jacobian, we find
\begin{equation}
\label{28apr16}
J_1=- \frac{4i (\bar\lambda_{\dot 1}^1 \lambda_1^1)^2}{\bar\lambda_{\dot 1}^2\lambda_1^2 (\bar\lambda_{\dot 1}^1 \lambda_1^1 +\bar\lambda_{\dot 2}^1 \lambda_2^1)}=\frac{2i \bar\lambda_{\dot 1}^1 \lambda_1^1}{(p^1)^0} = 2i \frac{[\mu 1]\langle \mu 1\rangle}{(p^1)^0}.
\end{equation}

Next, we give some intermediate steps of the computation that resulted in (\ref{4may19}).
To this end, we need to pass from $\delta(x_4)\delta(x_5)\delta(x_6)\delta(x_7)$ to $\delta(y_4)\delta(y_5)\delta(y_6)\delta(y_7)$, where
\begin{equation}
\label{16may7}
\begin{split}
x_4 &\equiv \langle 2\mu\rangle [\mu 1]-\langle 1\mu \rangle [\mu 2], \qquad x_5 \equiv \langle 1\mu\rangle [\mu 1]-\langle 2\mu \rangle [\mu 2],\\
x_6 &\equiv \langle 12\rangle ,\qquad x_7\equiv [12],\\
y_4&\equiv \lambda_1^1 - \lambda_2^1, \qquad y_5 \equiv \lambda_1^2 - \lambda_2^2,\qquad 
y_6\equiv \bar\lambda_1^{\dot 1} - \bar \lambda_2^{\dot 1}, \qquad y_7 \equiv \bar\lambda_1^{ \dot 2} - \bar \lambda_2^{ \dot 2}. 
\end{split}
\end{equation}
Computing explicitly, we find
\begin{equation}
\label{16may8}
\begin{split}
x_4 & = \mu^2 \bar\mu^{\dot 1}(\lambda_2^1 y_4 +\bar\lambda_2^{\dot 2}y_1)+\mu^2 \bar\mu^{\dot 2}(-\lambda_2^1 y_3 -\bar\lambda_{2}^{\dot 1}y_1)\\
&+ \mu^1 \bar\mu^{\dot 1}(-\lambda_2^2 y_4 -\bar\lambda_2^{\dot 2}y_2)+\mu^1 \bar\mu^{\dot 2}(\lambda_2^2 y_3 +\bar\lambda_2^{\dot 1}y_2),\\
x_5 &= \mu^2 \bar\mu^{\dot 1}(\lambda_1^1 y_4-\bar\lambda_2^{\dot 2}y_1)+\mu^2 \bar\mu^{\dot 2}(-\lambda_1^1 y_3 +\bar\lambda_{2}^{\dot 1}y_1)\\
&+ \mu^1 \bar\mu^{\dot 1}(-\lambda_1^2 y_4 +\bar\lambda_2^{\dot 2}y_2)+\mu^1 \bar\mu^{\dot 2}(\lambda_1^2 y_3 -\bar\lambda_2^{\dot 1}y_2),\\
x_6 &= \lambda_1^1 y_2 - \lambda_1^2 y_1, \qquad x_7 = -\bar\lambda_1^{\dot 1} y_4 + \bar\lambda_1^{\dot 2}y_3.
\end{split}
\end{equation}
Then, 
\begin{equation}
\label{16may9}
J_2\equiv \left|\frac{\partial (x_4,x_5,x_6,x_7)}{\partial (y_4,y_5,y_6,y_7)} \right|_{y_i=0}=[\mu 1][\mu 2] \langle \mu 1 \rangle (\langle \mu 1\rangle + \langle \mu 2\rangle)\Big|_{y_i=0}=
2[\mu 1]^2 \langle \mu1\rangle^2.
\end{equation}

Another computation that we would like to clarify is the passage from (\ref{9may12}) to (\ref{16jul12}), which implies that we need to convert $\delta(y_8)\delta(y_9)$ to $\delta(x_8)\delta(x_9)$ with
\begin{equation}
\begin{split}
\label{16jul9x1}
y_8 &\equiv [12]-[23], \qquad y_9\equiv [12]-[31]
\\
x_8 &\equiv \bar\lambda_1^{\dot 1}+\bar\lambda_2^{\dot 1}+\bar\lambda_3^{\dot 1}, \qquad x_9 \equiv \bar\lambda_1^{\dot 2}+\bar\lambda_2^{\dot 2}+\bar\lambda_3^{\dot 2}.
\end{split}
\end{equation}
It is straightforward to see that
\begin{equation}
\label{16may10}
y_8 = \bar\lambda_2^{\dot 2} x_8 -\bar\lambda_2^{\dot 1}x_9, \qquad y_9 = \bar\lambda_1^{\dot 1} x_9-\bar\lambda_1^{\dot 2}x_8,
\end{equation}
therefore,
\begin{equation}
\label{16may11}
J_3\equiv \left|\frac{\partial (y_8,y_9)}{\partial (x_8,x_9)} \right|_{x_i=0}= [12].
\end{equation}

Finally, we show how (\ref{16jul12}) can be rewritten as (\ref{16jul16}). To pass from $\delta(x_{10})\delta(x_{11})\delta(x_{12})\delta(x_{13})$
to $\delta(y_{10})\delta(y_{11})\delta(y_{12})\delta(y_{13})$ with 
\begin{equation}
\label{16may12}
\begin{split}
x_{10}&\equiv \lambda^{1}_1\bar\lambda^{\dot 1}_1+ \lambda^{1}_2\bar\lambda^{\dot 1}_2+\lambda^1_3\bar\lambda^{\dot 1}_3,\\
x_{11}&\equiv \lambda^{1}_1\bar\lambda^{\dot 2}_1+ \lambda^{1}_2\bar\lambda^{\dot 2}_2+\lambda^1_3\bar\lambda^{\dot 2}_3,\\
x_{12}&\equiv \lambda^{2}_1\bar\lambda^{\dot 1}_1+ \lambda^{2}_2\bar\lambda^{\dot 1}_2+\lambda^2_3\bar\lambda^{\dot 1}_3,\\
x_{13}&\equiv \lambda^{2}_1\bar\lambda^{\dot 2}_1+ \lambda^{2}_2\bar\lambda^{\dot 2}_2+\lambda^2_3\bar\lambda^{\dot 2}_3
\end{split}
\end{equation}
and
\begin{equation}
\label{16may13}
\begin{split}
y_{10}&\equiv\lambda_1^1 - \lambda_3^1, \qquad y_{11}\equiv \lambda_1^2-\lambda_3^2,\\
y_{12}&\equiv\lambda_2^1 - \lambda_3^1, \qquad y_{13}\equiv \lambda_2^2-\lambda_3^2
\end{split}
\end{equation}
we compute
\begin{equation}
\label{16may14}
\begin{split}
x_{10}&=\bar\lambda_1^{\dot 1} y_{10}+\bar\lambda_2^{\dot 1}y_{12}, \qquad x_{11}=\bar\lambda_1^{\dot 2} y_{10}+\bar\lambda_2^{\dot 2} y_{12},\\
x_{12}&=\bar\lambda_1^{\dot 1}y_{11}+ \bar\lambda_2^{\dot 1}y_{13}, \qquad x_{13}=\bar\lambda_1^{\dot 2} y_{11}+ \bar\lambda_2^{\dot 2}y_{13},
\end{split}
\end{equation}
where $x_8=0$ and $x_9=0$ was used to eliminate $\bar\lambda_3^{\dot 1}$ and $\bar\lambda_3^{\dot 2}$. We find that
\begin{equation}
\label{16may15}
J_4\equiv \left|\frac{\partial (x_{10},x_{11},x_{12},x_{13})}{\partial (y_{10},y_{11},y_{12},y_{13})} \right|_{y_i=0}=[12]^2.
\end{equation}

In the above manipulations one needs to know how the standard formula
\begin{equation}
\label{5aug10}
\delta(\alpha x) = \frac{1}{|\alpha|}\delta(x)
\end{equation}
for real $\alpha$ and $x$ can be extended to complex variables. This issue is quite common in higher-spin theories, but the precise phases entering this formula are usually irrelevant due to the fact that one mostly deals with parity-invariant theories and the problematic phases cancel out. In the chiral case manipulations with the arguments of delta functions were discussed in appendix B.4 of \cite{Iazeolla:2011cb}. Above we ignored such phase factors. This can be justified e.g. by going to the $(2,2)$ signature -- for which $\lambda$ and $\bar\lambda$ are real, therefore, (\ref{5aug10}) can be applied -- and then analytically continuing the result to the Lorentzian signature. Unwanted overall signs, if needed, can be removed by field redefinitions and by redefinitions of the  coupling constant $g$.

\section{Poincare subalgebra}
\label{app:poin}

In this section we consider commutator of transformations (\ref{11may2})
\begin{equation}
\label{6may1xxx3}
\begin{split}
[\Phi,\varepsilon]_{\ltimes}(\lambda,\bar\lambda)
= \int d^2\bar\lambda_2
(e^{[\lambda 2]}-e^{-[\lambda 2]})
\Phi(\lambda,\bar\lambda - \bar\lambda_2)
\varepsilon(\lambda,\bar\lambda_2)
\end{split}
\end{equation}
with gauge parameters of the form
\begin{equation}
\label{6may1xxx4}
\varepsilon= (\lambda_\alpha)^m\left(\frac{\partial}{\partial\bar\lambda^{\dot\alpha}} \right)^n \delta^2(\bar\lambda).
\end{equation}
Parameters of this form become polynomials after the Fourier transform in $\bar\lambda$. Plugging (\ref{6may1xxx4}) and integrating by parts, we find 
\begin{equation}
\label{6may1xxx5}
\begin{split}
[\Phi,\varepsilon]_{\ltimes}(\lambda,\bar\lambda)
=
 (\lambda_\alpha)^m
 (-1)^n
\left(\frac{\partial}{\partial\bar\lambda_2^{\dot\alpha}} \right)^n \Big[
(e^{[\lambda 2]}-e^{-[\lambda 2]})
\Phi(\lambda,\bar\lambda - \bar\lambda_2)
\Big]_{\bar\lambda_2=0}.
\end{split}
\end{equation}
This is evaluated using the standard Leibniz rule and gives
\begin{equation}
\label{6may1xxx6}
\begin{split}
[\Phi,\varepsilon]_{\ltimes}(\lambda,\bar\lambda)
=
 (-1)^n (\lambda_\alpha)^m
\sum_{i=0}^n
\frac{n!(-1)^i}{i!(n-i)!}\big(1-(-1)^{n-i}\big)
\left(\bar\lambda_{\dot\alpha}\right)^{n-i}
\left(\frac{\partial}{\partial\bar\lambda^{\dot\alpha}} \right)^n 
\Phi(\lambda,\bar\lambda).
\end{split}
\end{equation}
Note that this expression is only non-vanishing for $n-i$ odd.

Focusing on the case $m=1$ and $n=1$, we find
\begin{equation}
\label{app16may1}
\left[\Phi, \lambda_\alpha \frac{\partial}{\partial\bar\lambda^{\dot\alpha}}  \delta^2(\bar\lambda)\right] = -2\lambda_\alpha \bar\lambda_{\dot\alpha}\Phi,
\end{equation}
which, up to a factor, gives translations. Analogously, for $m=0$ and $n=2$ we find
\begin{equation}
\label{app16may2}
\left[\Phi, \left(\frac{\partial}{\partial\bar\lambda^{\dot\alpha}} \right)^2 \delta^2(\bar\lambda)\right] = -4\bar\lambda_{\dot\alpha} \frac{\partial}{\partial \bar\lambda^{\dot\alpha}}\Phi,
\end{equation}
which up to a factor reproduces the $\bar J$ part of the Lorentz transformations.

Clearly, in a similar manner we can generate translations and $J$ from the $\rtimes$-commutator  of opposite chirality.

\bibliography{hsalgebra}

\providecommand{\href}[2]{#2}\begingroup\raggedright\begin{thebibliography}{10}

\bibitem{Weinberg:1964ew}
S.~Weinberg, \emph{{Photons and Gravitons in $S$-Matrix Theory: Derivation of
  Charge Conservation and Equality of Gravitational and Inertial Mass}},
  \href{https://doi.org/10.1103/PhysRev.135.B1049}{\emph{Phys. Rev.} {\bfseries
  135} (1964) B1049}.

\bibitem{Coleman:1967ad}
S.R.~Coleman and J.~Mandula, \emph{{All Possible Symmetries of the S Matrix}},
  \href{https://doi.org/10.1103/PhysRev.159.1251}{\emph{Phys. Rev.} {\bfseries
  159} (1967) 1251}.

\bibitem{Bekaert:2010hw}
X.~Bekaert, N.~Boulanger and P.~Sundell, \emph{{How higher-spin gravity
  surpasses the spin two barrier: no-go theorems versus yes-go examples}},
  \href{https://doi.org/10.1103/RevModPhys.84.987}{\emph{Rev. Mod. Phys.}
  {\bfseries 84} (2012) 987} [\href{https://arxiv.org/abs/1007.0435}{{\ttfamily
  1007.0435}}].

\bibitem{Metsaev:1991mt}
R.R.~Metsaev, \emph{{Poincare invariant dynamics of massless higher spins:
  Fourth order analysis on mass shell}},
  \href{https://doi.org/10.1142/S0217732391000348}{\emph{Mod. Phys. Lett. A}
  {\bfseries 6} (1991) 359}.

\bibitem{Metsaev:1991nb}
R.R.~Metsaev, \emph{{S matrix approach to massless higher spins theory. 2: The
  Case of internal symmetry}},
  \href{https://doi.org/10.1142/S0217732391002839}{\emph{Mod. Phys. Lett. A}
  {\bfseries 6} (1991) 2411}.

\bibitem{Ponomarev:2016lrm}
D.~Ponomarev and E.D.~Skvortsov, \emph{{Light-Front Higher-Spin Theories in
  Flat Space}}, \href{https://doi.org/10.1088/1751-8121/aa56e7}{\emph{J. Phys.
  A} {\bfseries 50} (2017) 095401}
  [\href{https://arxiv.org/abs/1609.04655}{{\ttfamily 1609.04655}}].

\bibitem{Ponomarev:2017nrr}
D.~Ponomarev, \emph{{Chiral Higher Spin Theories and Self-Duality}},
  \href{https://doi.org/10.1007/JHEP12(2017)141}{\emph{JHEP} {\bfseries 12}
  (2017) 141} [\href{https://arxiv.org/abs/1710.00270}{{\ttfamily
  1710.00270}}].

\bibitem{Skvortsov:2018jea}
E.D.~Skvortsov, T.~Tran and M.~Tsulaia, \emph{{Quantum Chiral Higher Spin
  Gravity}}, \href{https://doi.org/10.1103/PhysRevLett.121.031601}{\emph{Phys.
  Rev. Lett.} {\bfseries 121} (2018) 031601}
  [\href{https://arxiv.org/abs/1805.00048}{{\ttfamily 1805.00048}}].

\bibitem{Skvortsov:2020wtf}
E.~Skvortsov, T.~Tran and M.~Tsulaia, \emph{{More on Quantum Chiral Higher Spin
  Gravity}}, \href{https://doi.org/10.1103/PhysRevD.101.106001}{\emph{Phys.
  Rev. D} {\bfseries 101} (2020) 106001}
  [\href{https://arxiv.org/abs/2002.08487}{{\ttfamily 2002.08487}}].

\bibitem{Skvortsov:2020gpn}
E.~Skvortsov and T.~Tran, \emph{{One-loop Finiteness of Chiral Higher Spin
  Gravity}}, \href{https://doi.org/10.1007/JHEP07(2020)021}{\emph{JHEP}
  {\bfseries 07} (2020) 021}
  [\href{https://arxiv.org/abs/2004.10797}{{\ttfamily 2004.10797}}].

\bibitem{Sezgin:2002rt}
E.~Sezgin and P.~Sundell, \emph{{Massless higher spins and holography}},
  \href{https://doi.org/10.1016/S0550-3213(02)00739-3}{\emph{Nucl. Phys. B}
  {\bfseries 644} (2002) 303}
  [\href{https://arxiv.org/abs/hep-th/0205131}{{\ttfamily hep-th/0205131}}].

\bibitem{Klebanov:2002ja}
I.R.~Klebanov and A.M.~Polyakov, \emph{{AdS dual of the critical O(N) vector
  model}}, \href{https://doi.org/10.1016/S0370-2693(02)02980-5}{\emph{Phys.
  Lett. B} {\bfseries 550} (2002) 213}
  [\href{https://arxiv.org/abs/hep-th/0210114}{{\ttfamily hep-th/0210114}}].

\bibitem{Maldacena:2011jn}
J.~Maldacena and A.~Zhiboedov, \emph{{Constraining Conformal Field Theories
  with A Higher Spin Symmetry}},
  \href{https://doi.org/10.1088/1751-8113/46/21/214011}{\emph{J. Phys. A}
  {\bfseries 46} (2013) 214011}
  [\href{https://arxiv.org/abs/1112.1016}{{\ttfamily 1112.1016}}].

\bibitem{Boulanger:2013zza}
N.~Boulanger, D.~Ponomarev, E.D.~Skvortsov and M.~Taronna, \emph{{On the
  uniqueness of higher-spin symmetries in AdS and CFT}},
  \href{https://doi.org/10.1142/S0217751X13501625}{\emph{Int. J. Mod. Phys. A}
  {\bfseries 28} (2013) 1350162}
  [\href{https://arxiv.org/abs/1305.5180}{{\ttfamily 1305.5180}}].

\bibitem{Alba:2013yda}
V.~Alba and K.~Diab, \emph{{Constraining conformal field theories with a higher
  spin symmetry in d=4}},  \href{https://arxiv.org/abs/1307.8092}{{\ttfamily
  1307.8092}}.

\bibitem{Taronna:2016ats}
M.~Taronna, \emph{{Pseudo-local Theories: A Functional Class Proposal}},  in
  \emph{{International Workshop on Higher Spin Gauge Theories}}, pp.~59--84,
  2017, \href{https://doi.org/10.1142/9789813144101_0006}{DOI}
  [\href{https://arxiv.org/abs/1602.08566}{{\ttfamily 1602.08566}}].

\bibitem{Bekaert:2016ezc}
X.~Bekaert, J.~Erdmenger, D.~Ponomarev and C.~Sleight, \emph{{Bulk quartic
  vertices from boundary four-point correlators}},  in \emph{{International
  Workshop on Higher Spin Gauge Theories}}, pp.~291--303, 2017,
  \href{https://doi.org/10.1142/9789813144101_0015}{DOI}
  [\href{https://arxiv.org/abs/1602.08570}{{\ttfamily 1602.08570}}].

\bibitem{Fradkin:1986ka}
E.S.~Fradkin and M.A.~Vasiliev, \emph{{Candidate to the Role of Higher Spin
  Symmetry}}, \href{https://doi.org/10.1016/S0003-4916(87)80025-8}{\emph{Annals
  Phys.} {\bfseries 177} (1987) 63}.

\bibitem{Eastwood:2002su}
M.G.~Eastwood, \emph{{Higher symmetries of the Laplacian}},
  \href{https://doi.org/10.4007/annals.2005.161.1645}{\emph{Annals Math.}
  {\bfseries 161} (2005) 1645}
  [\href{https://arxiv.org/abs/hep-th/0206233}{{\ttfamily hep-th/0206233}}].

\bibitem{Vasiliev:2003ev}
M.A.~Vasiliev, \emph{{Nonlinear equations for symmetric massless higher spin
  fields in (A)dS(d)}},
  \href{https://doi.org/10.1016/S0370-2693(03)00872-4}{\emph{Phys. Lett. B}
  {\bfseries 567} (2003) 139}
  [\href{https://arxiv.org/abs/hep-th/0304049}{{\ttfamily hep-th/0304049}}].

\bibitem{Vasiliev:1986qx}
M.A.~Vasiliev, \emph{{Extended Higher Spin Superalgebras and Their Realizations
  in Terms of Quantum Operators}}, {\emph{Fortsch. Phys.} {\bfseries 36} (1988)
  33}.

\bibitem{Colombo:2012jx}
N.~Colombo and P.~Sundell, \emph{{Higher Spin Gravity Amplitudes From Zero-form
  Charges}},  \href{https://arxiv.org/abs/1208.3880}{{\ttfamily 1208.3880}}.

\bibitem{Didenko:2012tv}
V.E.~Didenko and E.D.~Skvortsov, \emph{{Exact higher-spin symmetry in CFT: all
  correlators in unbroken Vasiliev theory}},
  \href{https://doi.org/10.1007/JHEP04(2013)158}{\emph{JHEP} {\bfseries 04}
  (2013) 158} [\href{https://arxiv.org/abs/1210.7963}{{\ttfamily 1210.7963}}].

\bibitem{Gelfond:2013xt}
O.A.~Gelfond and M.A.~Vasiliev, \emph{{Operator algebra of free conformal
  currents via twistors}},
  \href{https://doi.org/10.1016/j.nuclphysb.2013.09.001}{\emph{Nucl. Phys. B}
  {\bfseries 876} (2013) 871}
  [\href{https://arxiv.org/abs/1301.3123}{{\ttfamily 1301.3123}}].

\bibitem{Didenko:2013bj}
V.E.~Didenko, J.~Mei and E.D.~Skvortsov, \emph{{Exact higher-spin symmetry in
  CFT: free fermion correlators from Vasiliev Theory}},
  \href{https://doi.org/10.1103/PhysRevD.88.046011}{\emph{Phys. Rev. D}
  {\bfseries 88} (2013) 046011}
  [\href{https://arxiv.org/abs/1301.4166}{{\ttfamily 1301.4166}}].

\bibitem{Maldacena:2012sf}
J.~Maldacena and A.~Zhiboedov, \emph{{Constraining conformal field theories
  with a slightly broken higher spin symmetry}},
  \href{https://doi.org/10.1088/0264-9381/30/10/104003}{\emph{Class. Quant.
  Grav.} {\bfseries 30} (2013) 104003}
  [\href{https://arxiv.org/abs/1204.3882}{{\ttfamily 1204.3882}}].

\bibitem{Gerasimenko:2021sxj}
P.~Gerasimenko, A.~Sharapov and E.~Skvortsov, \emph{{Slightly broken higher
  spin symmetry: general structure of correlators}},
  \href{https://doi.org/10.1007/JHEP01(2022)097}{\emph{JHEP} {\bfseries 01}
  (2022) 097} [\href{https://arxiv.org/abs/2108.05441}{{\ttfamily
  2108.05441}}].

\bibitem{Elvang:2013cua}
H.~Elvang and Y.-t.~Huang, \emph{{Scattering Amplitudes}},
  \href{https://arxiv.org/abs/1308.1697}{{\ttfamily 1308.1697}}.

\bibitem{Bolotin:1999fa}
K.I.~Bolotin and M.A.~Vasiliev, \emph{{Star product and massless free field
  dynamics in AdS(4)}},
  \href{https://doi.org/10.1016/S0370-2693(00)00307-5}{\emph{Phys. Lett. B}
  {\bfseries 479} (2000) 421}
  [\href{https://arxiv.org/abs/hep-th/0001031}{{\ttfamily hep-th/0001031}}].

\bibitem{Nagaraj:2018nxq}
B.~Nagaraj and D.~Ponomarev, \emph{{Spinor-Helicity Formalism for Massless
  Fields in AdS$_4$}},
  \href{https://doi.org/10.1103/PhysRevLett.122.101602}{\emph{Phys. Rev. Lett.}
  {\bfseries 122} (2019) 101602}
  [\href{https://arxiv.org/abs/1811.08438}{{\ttfamily 1811.08438}}].

\bibitem{Nagaraj:2019zmk}
B.~Nagaraj and D.~Ponomarev, \emph{{Spinor-helicity formalism for massless
  fields in AdS$_{4}$. Part II. Potentials}},
  \href{https://doi.org/10.1007/JHEP06(2020)068}{\emph{JHEP} {\bfseries 06}
  (2020) 068} [\href{https://arxiv.org/abs/1912.07494}{{\ttfamily
  1912.07494}}].

\bibitem{Nagaraj:2020sji}
B.~Nagaraj and D.~Ponomarev, \emph{{Spinor-helicity formalism for massless
  fields in AdS$_{4}$ III: contact four-point amplitudes}},
  \href{https://doi.org/10.1007/JHEP08(2020)012}{\emph{JHEP} {\bfseries 08}
  (2020) 012} [\href{https://arxiv.org/abs/2004.07989}{{\ttfamily
  2004.07989}}].

\bibitem{Joung:2014qya}
E.~Joung and K.~Mkrtchyan, \emph{{Notes on higher-spin algebras: minimal
  representations and structure constants}},
  \href{https://doi.org/10.1007/JHEP05(2014)103}{\emph{JHEP} {\bfseries 05}
  (2014) 103} [\href{https://arxiv.org/abs/1401.7977}{{\ttfamily 1401.7977}}].

\bibitem{Krasnov:2021nsq}
K.~Krasnov, E.~Skvortsov and T.~Tran, \emph{{Actions for self-dual Higher Spin
  Gravities}}, \href{https://doi.org/10.1007/JHEP08(2021)076}{\emph{JHEP}
  {\bfseries 08} (2021) 076}
  [\href{https://arxiv.org/abs/2105.12782}{{\ttfamily 2105.12782}}].

\bibitem{Krasnov:2021cva}
K.~Krasnov and E.~Skvortsov, \emph{{Flat self-dual gravity}},
  \href{https://doi.org/10.1007/JHEP08(2021)082}{\emph{JHEP} {\bfseries 08}
  (2021) 082} [\href{https://arxiv.org/abs/2106.01397}{{\ttfamily
  2106.01397}}].

\bibitem{Skvortsov:2022syz}
E.~Skvortsov and R.~Van~Dongen, \emph{{Minimal models of field theories: Chiral
  Higher Spin Gravity}},  \href{https://arxiv.org/abs/2204.10285}{{\ttfamily
  2204.10285}}.

\bibitem{Sharapov:2022dfw}
A.~Sharapov, E.~Skvortsov, A.~Sukhanov and R.~Van~Dongen, \emph{{Minimal model
  of Chiral Higher Spin Gravity}},
  \href{https://arxiv.org/abs/2205.07794}{{\ttfamily 2205.07794}}.

\bibitem{VOA1}
J.~Lepowsky and H.~Li, \emph{{Introduction to Vertex Operator Algebras and
  Their Representations}}, Progress in Mathematics, {Birkh{\"a}user Boston,
  MA}, 1~ed. (2004).

\bibitem{Nozaradan:2008zq}
C.~Nozaradan, \emph{{Introduction to Vertex Algebras}},
  \href{https://arxiv.org/abs/0809.1380}{{\ttfamily 0809.1380}}.

\bibitem{Bern:2008qj}
Z.~Bern, J.J.M.~Carrasco and H.~Johansson, \emph{{New Relations for
  Gauge-Theory Amplitudes}},
  \href{https://doi.org/10.1103/PhysRevD.78.085011}{\emph{Phys. Rev. D}
  {\bfseries 78} (2008) 085011}
  [\href{https://arxiv.org/abs/0805.3993}{{\ttfamily 0805.3993}}].

\bibitem{Monteiro:2011pc}
R.~Monteiro and D.~O'Connell, \emph{{The Kinematic Algebra From the Self-Dual
  Sector}}, \href{https://doi.org/10.1007/JHEP07(2011)007}{\emph{JHEP}
  {\bfseries 07} (2011) 007} [\href{https://arxiv.org/abs/1105.2565}{{\ttfamily
  1105.2565}}].

\bibitem{Chalmers:1996rq}
G.~Chalmers and W.~Siegel, \emph{{The Selfdual sector of QCD amplitudes}},
  \href{https://doi.org/10.1103/PhysRevD.54.7628}{\emph{Phys. Rev. D}
  {\bfseries 54} (1996) 7628}
  [\href{https://arxiv.org/abs/hep-th/9606061}{{\ttfamily hep-th/9606061}}].

\bibitem{Vasiliev:1999ba}
M.A.~Vasiliev, \emph{{Higher spin gauge theories: Star product and AdS space}},
   \href{https://arxiv.org/abs/hep-th/9910096}{{\ttfamily hep-th/9910096}}.

\bibitem{Didenko:2009td}
V.E.~Didenko and M.A.~Vasiliev, \emph{{Static BPS black hole in 4d higher-spin
  gauge theory}},
  \href{https://doi.org/10.1016/j.physletb.2009.11.023}{\emph{Phys. Lett. B}
  {\bfseries 682} (2009) 305}
  [\href{https://arxiv.org/abs/0906.3898}{{\ttfamily 0906.3898}}].

\bibitem{Fradkin:1987ks}
E.S.~Fradkin and M.A.~Vasiliev, \emph{{On the Gravitational Interaction of
  Massless Higher Spin Fields}},
  \href{https://doi.org/10.1016/0370-2693(87)91275-5}{\emph{Phys. Lett. B}
  {\bfseries 189} (1987) 89}.

\bibitem{Boulanger:2008tg}
N.~Boulanger, S.~Leclercq and P.~Sundell, \emph{{On The Uniqueness of Minimal
  Coupling in Higher-Spin Gauge Theory}},
  \href{https://doi.org/10.1088/1126-6708/2008/08/056}{\emph{JHEP} {\bfseries
  08} (2008) 056} [\href{https://arxiv.org/abs/0805.2764}{{\ttfamily
  0805.2764}}].

\bibitem{Vasiliev:1986td}
M.A.~Vasiliev, \emph{{Free Massless Fields of Arbitrary Spin in the De Sitter
  Space and Initial Data for a Higher Spin Superalgebra}}, {\emph{Fortsch.
  Phys.} {\bfseries 35} (1987) 741}.

\bibitem{Bekaert:2005ka}
X.~Bekaert and N.~Boulanger, \emph{{Gauge invariants and Killing tensors in
  higher-spin gauge theories}},
  \href{https://doi.org/10.1016/j.nuclphysb.2005.06.009}{\emph{Nucl. Phys. B}
  {\bfseries 722} (2005) 225}
  [\href{https://arxiv.org/abs/hep-th/0505068}{{\ttfamily hep-th/0505068}}].

\bibitem{Weinberg}
{Weinberg, Steven}, \emph{{The Quantum Theory of Fields}}, vol.~{1,2,3},
  {Cambridge University Press} ({1995}).

\bibitem{Joung:2015eny}
E.~Joung, S.~Nakach and A.A.~Tseytlin, \emph{{Scalar scattering via conformal
  higher spin exchange}},
  \href{https://doi.org/10.1007/JHEP02(2016)125}{\emph{JHEP} {\bfseries 02}
  (2016) 125} [\href{https://arxiv.org/abs/1512.08896}{{\ttfamily
  1512.08896}}].

\bibitem{Beccaria:2016syk}
M.~Beccaria, S.~Nakach and A.A.~Tseytlin, \emph{{On triviality of S-matrix in
  conformal higher spin theory}},
  \href{https://doi.org/10.1007/JHEP09(2016)034}{\emph{JHEP} {\bfseries 09}
  (2016) 034} [\href{https://arxiv.org/abs/1607.06379}{{\ttfamily
  1607.06379}}].

\bibitem{Sleight:2016xqq}
C.~Sleight and M.~Taronna, \emph{{Higher-Spin Algebras, Holography and Flat
  Space}}, \href{https://doi.org/10.1007/JHEP02(2017)095}{\emph{JHEP}
  {\bfseries 02} (2017) 095}
  [\href{https://arxiv.org/abs/1609.00991}{{\ttfamily 1609.00991}}].

\bibitem{Ponomarev:2016cwi}
D.~Ponomarev, \emph{{Off-Shell Spinor-Helicity Amplitudes from Light-Cone
  Deformation Procedure}},
  \href{https://doi.org/10.1007/JHEP12(2016)117}{\emph{JHEP} {\bfseries 12}
  (2016) 117} [\href{https://arxiv.org/abs/1611.00361}{{\ttfamily
  1611.00361}}].

\bibitem{Penedones:2010ue}
J.~Penedones, \emph{{Writing CFT correlation functions as AdS scattering
  amplitudes}}, \href{https://doi.org/10.1007/JHEP03(2011)025}{\emph{JHEP}
  {\bfseries 03} (2011) 025} [\href{https://arxiv.org/abs/1011.1485}{{\ttfamily
  1011.1485}}].

\bibitem{Fitzpatrick:2011dm}
A.L.~Fitzpatrick and J.~Kaplan, \emph{{Unitarity and the Holographic
  S-Matrix}}, \href{https://doi.org/10.1007/JHEP10(2012)032}{\emph{JHEP}
  {\bfseries 10} (2012) 032} [\href{https://arxiv.org/abs/1112.4845}{{\ttfamily
  1112.4845}}].

\bibitem{Ponomarev:2019ofr}
D.~Ponomarev, \emph{{From bulk loops to boundary large-N expansion}},
  \href{https://doi.org/10.1007/JHEP01(2020)154}{\emph{JHEP} {\bfseries 01}
  (2020) 154} [\href{https://arxiv.org/abs/1908.03974}{{\ttfamily
  1908.03974}}].

\bibitem{Meltzer:2019nbs}
D.~Meltzer, E.~Perlmutter and A.~Sivaramakrishnan, \emph{{Unitarity Methods in
  AdS/CFT}}, \href{https://doi.org/10.1007/JHEP03(2020)061}{\emph{JHEP}
  {\bfseries 03} (2020) 061}
  [\href{https://arxiv.org/abs/1912.09521}{{\ttfamily 1912.09521}}].

\bibitem{Campoleoni:2021blr}
A.~Campoleoni and S.~Pekar, \emph{{Carrollian and Galilean conformal
  higher-spin algebras in any dimensions}},
  \href{https://doi.org/10.1007/JHEP02(2022)150}{\emph{JHEP} {\bfseries 02}
  (2022) 150} [\href{https://arxiv.org/abs/2110.07794}{{\ttfamily
  2110.07794}}].

\bibitem{Ponomarev:2021xdq}
D.~Ponomarev, \emph{{3d conformal fields with manifest sl(2,
  \ensuremath{\mathbb{C}})}},
  \href{https://doi.org/10.1007/JHEP06(2021)055}{\emph{JHEP} {\bfseries 06}
  (2021) 055} [\href{https://arxiv.org/abs/2104.02770}{{\ttfamily
  2104.02770}}].

\bibitem{Iazeolla:2011cb}
C.~Iazeolla and P.~Sundell, \emph{{Families of exact solutions to Vasiliev's 4D
  equations with spherical, cylindrical and biaxial symmetry}},
  \href{https://doi.org/10.1007/JHEP12(2011)084}{\emph{JHEP} {\bfseries 12}
  (2011) 084} [\href{https://arxiv.org/abs/1107.1217}{{\ttfamily 1107.1217}}].

\end{thebibliography}\endgroup
\bibliographystyle{JHEP}

\end{document}